\documentclass{jpp}
\usepackage[colorlinks=true,linkcolor=blue,citecolor=blue]{hyperref}
\usepackage{graphicx}
\usepackage{bm}
\usepackage[utf8]{inputenc}
\usepackage[T1]{fontenc}
\usepackage{amsmath}
\usepackage{dsfont}
\usepackage[section]{placeins}
\usepackage{multicol}
\usepackage{hhline}
\usepackage{tikz}
\usepackage{booktabs}

\let\originalleft\left
\let\originalright\right
\renewcommand{\left}{\mathopen{}\mathclose\bgroup\originalleft}
\renewcommand{\right}{\aftergroup\egroup\originalright}

\newcommand{\llbrace}{\left\lbrace}
\newcommand{\rrbrace}{\right\rbrace}

\newcommand*\circled[1]{\tikz[baseline=(char.base)]{
            \node[shape=circle,draw,inner sep=1pt] (char) {#1};}}
\newcommand{\sd}{\mathrm{d}}
\newcommand{\grad}{\nabla}
\newcommand{\gradperp}{\nabla_{\perp}}

\newcommand{\unit}[1]{\hat{\bm{#1}}}
\newcommand{\kperp}{k_{\perp}}
\newcommand{\kpar}{k_{\parallel}}
\newcommand{\vchi}{\bm{v}_{\chi}}
\newcommand{\vpar}{v_{\parallel}}
\newcommand{\vperp}{v_{\perp}}
\newcommand{\vthi}{v_{\mathrm{th}i}}

\newcommand{\vths}{v_{\mathrm{th}s}}

\newcommand{\aperp}{\bm{A}_{\perp}}
\newcommand{\apar}{A_{\parallel}}
\newcommand{\bpar}{B_{\parallel}}

\newcommand{\avg}[2]{\left<#1\right>_{#2}}

\newcommand{\gyror}[1]{\left<#1\right>_{\bm{r}}}
\newcommand{\gyroRs}[1]{\left<#1\right>_{\bm{R}_s}}

\shorttitle{Zonal flows and saturation of electromagnetic turbulence}
\shortauthor{Y. Zhang et al.}

\title{Zonal-flow generation and saturation of electromagnetic ion-scale turbulence in tokamaks}

\author{Y.~Zhang\aff{1,2,3,4}
\corresp{\email{yujia.zhang@ukaea.uk}},
T.~Adkins\aff{5},
M.~Barnes\aff{1,6},
A.~V.~Dudkovskaia\aff{3},
M.~R.~Hardman\aff{4},
P.~G.~Ivanov\aff{4},
D.~Kennedy\aff{4},
A.~A.~Schekochihin\aff{1,7}
}

\affiliation{\aff{1}Rudolf Peierls Centre for Theoretical Physics, University of Oxford, Oxford OX1 3PU, UK
\aff{2}St Edmund Hall, Oxford OX1 4AR, UK
\aff{3}Tokamak Energy Ltd, 173 Brook Dr, Milton, Abingdon OX14 4SD
\aff{4}United Kingdom Atomic Energy Authority, Culham Campus, Abingdon, Oxfordshire, OX14 3DB, UK
\aff{5}Princeton Plasma Physics Laboratory, 100 Stellarator Road, Princeton, NJ 08540
\aff{6}University College, Oxford OX1 4BH, UK
\aff{7}Merton College, Oxford OX1 4JD, UK}

\begin{document}

\maketitle

\begin{abstract}
Local flux-tube gyrokinetic simulations of ion-scale turbulence in tokamak plasmas at finite plasma beta are conducted to investigate the generation of zonal flows via turbulent stresses. A parameter scan in the safety factor $q$ and electron beta $\beta_e$ reveals a transition from low- to high-transport states when $\beta_{\mathrm{eff}} \equiv q^2\beta_e$ exceeds a certain critical value $C_{\mathrm{nl}}$. While the linear stability limits for kinetic and ideal ballooning modes also scale as $\beta_e \propto 1/q^2$, they lie above the observed transition, indicating that the effect is not due to linear instabilities but to nonlinear dynamics. At low $\beta_{\mathrm{eff}}$, Reynolds stress dominates and drives zonal flows. At higher values, Maxwell stress becomes comparable, suppressing zonal-flow formation and leading to divergent transport. This nonlinear-transition boundary is determined for both the Cyclone Base Case and a spherical tokamak (ST40) configuration, suggesting that the relation $\beta_{\mathrm{eff}} = C_{\mathrm{nl}}$ may have broader applicability, though $C_{\mathrm{nl}}$ appears to be configuration-dependent. For the Cyclone Base Case, the ratio of energy transfer rates into zonal flows due to Maxwell and Reynolds stresses is observed empirically to scale as $\beta_e$ for $\beta_e$ below a critical value $\beta_{e,\mathrm{sb}}$ (scaling breakdown). The value of $\beta_{e,\mathrm{sb}}$ is found to increase with decreasing aspect ratio, suggesting that the linear scaling remains valid over a wider range of $\beta_e$ for more compact magnetic equilibria. This low-$\beta_e$ scaling provides the basis for a practical method to predict the nonlinear-transition threshold with minimal reliance on highly electromagnetic nonlinear simulations.
\end{abstract}

\section{Introduction}
\label{sec:intro}

In tokamak plasmas, energy confinement is often limited by turbulent transport driven by a variety of microinstabilities. Among the most important of these are ion-temperature-gradient (ITG) instabilities \citep{horton_1981, waltz_1988, romanelli_1989, Cowley_1991, Kotschenreuther_1995a}. ITG instabilities are electrostatic in nature, meaning that they are primarily driven by the ion temperature gradient and can exist in the absence of magnetic fluctuations. The ITG instabilities then give rise to ITG turbulence, leading to heat transport. Fortunately, ITG turbulence is typically regulated by the self-generation of zonal flows \citep{lin_1998, dimits_2000, roger_2000}. These zonal flows shear structures to dissipative scales, and thus limit the transport to levels that are observed in experiments.

However, for spherical tokamaks and certain reactor-relevant plasmas, the value of plasma beta is large enough that the magnetic fluctuations play an important role and can no longer be neglected. First-principles simulations of these plasmas indicate that large enough beta leads to extreme levels of transport or even lack of saturation. For instance, the turbulence driven by electromagnetic instabilities like kinetic ballooning modes (KBMs) \citep{Tang_1980, Aleynikova_2017, Adkins_2022} has been observed in the simulations to cause large transport even when the KBM growth rates are subdominant to electrostatic instabilities \citep{Mulholland_2024}. There are also cases where the linear instability exhibits characteristics of both electrostatic and electromagnetic modes \citep{Kennedy_2023,Giacomin_2024}, leading to transport levels significantly higher than those consistent with the power balance.

It is not clear whether the presence of electromagnetic instabilities is solely responsible for the observed high transport levels in the simulations. As found in a number of studies, the onset of electromagnetic instabilities does not necessarily imply such transport: gyrokinetic simulations of KBM-driven turbulence can achieve well-saturated turbulence \citep{MJ_2008, Maeyama_2014, Kumar_2021, Najlaoui_2025}, and can even produce heat fluxes considerably lower than the corresponding electrostatic limit \citep{McKinney_2021}. Conversely, in the absence of purely electromagnetic instabilities, the turbulence may still enter a regime of `high-beta runaway', characterized by extreme transport levels \citep{Waltz_2010, MJ_2013_finitebeta_NZT, Rath_2022, Giacomin_2024, zhang_2026}. This runaway behaviour of ion-scale turbulence is associated with the emergence of streamers -- radially-elongated, coherent structures -- that efficiently transport heat across the system, often once the plasma beta exceeds some critical value. Such a transition is often referred to as a `non-zonal' transition \citep{MJ_2013_finitebeta_NZT} to reflect both the corresponding suppression of zonal flow and to distinguish it from the electromagnetic linear-stability threshold. In this work, we shall be using the term `EM high-transport state' to describe a state with extremely large or divergent turbulent transport when exceeding critical values of beta. 

Several explanations have been proposed for the non-zonal transition. \cite{Waltz_2010} posited that the zonal structures are unstable to tertiary excitation of electromagnetic instabilities, such as KBMs, and that these tertiary modes then grow to dominate over the zonal modes. However, it was later shown that the zonal amplitudes generated by turbulence are insufficient to excite KBMs under typical conditions \citep{MJ_2013_instab}. Instead, \cite{MJ_2013_finitebeta_NZT, MJ_2014} argued that the transition arose from increased magnetic stochasticity at higher plasma beta, accompanied by fast electron streaming along the radially perturbed field lines, effectively `short-circuiting' the zonal structures. The presence of magnetic stochasticity -- and its quadratic scaling with the plasma beta in electromagnetic ITG turbulence -- had been discussed previously \citep{Nevins_2011} in contexts separate from the non-zonal transition. \cite{MJ_2013_finitebeta_NZT} proposed a new physical insight: while the tearing-parity magnetic perturbations dominate magnetic stochasticity below the non-zonal threshold, twisting-parity magnetic perturbations begin to contribute significantly at the threshold, leading to a discontinuous elevation of the magnetic stochasticity level. This change in twisting-parity behaviour was attributed to the radial excursions of field lines exceeding the turbulence’s radial correlation length. Consequently, it was argued that, above the threshold, stochasticity might exhibit a stronger scaling with magnetic energy due to contributions from modes of both parities. However, a more recent work \citep{Rath_2022} found no evidence for this stronger stochasticity scaling with magnetic energy above the non-zonal transition threshold. 

An alternative explanation proposed in \cite{Rath_2022} is that, at the non-zonal transition threshold, the Reynolds stress -- which normally drives zonal flows -- is completely cancelled by the Maxwell stress. The argument for thinking in terms of competing stresses is physically intuitive: if the non-zonal states result from weakened zonal flows, then directly analysing the evolution of zonal flows should reveal the underlying physics, where turbulent stresses emerge naturally as the relevant quantities. This kind of argument was used by \cite{ivanov_2020, ivanov_2022} to propose a simple picture of the \cite{dimits_2000} transition from low- to high-transport states in electrostatic turbulence: in a fluid model of such turbulence, the competition is between Reynolds and diamagnetic stresses, and the transition occurs when the latter overcome the former, which happens above a certain critical value of the temperature gradient. In a recent extension of this fluid model to electromagnetic turbulence, \cite{zhang_2026} showed that Maxwell stress acted in alliance with the diamagnetic stress, winning the competition with Reynolds stress at an ever lower critical value of the temperature gradient as plasma beta was increased. Thus, our approach is to conceptualise the `high-beta runaway' as a beta dependence of the Dimits-transition threshold. In this paper, we extend this approach from a $Z$-pinch fluid model to a tokamak-geometry, gyrokinetic one and show that the evolution of zonal flow is controlled by the competition between a combination of linear and nonlinear stresses. 

The remainder of this paper is organised as follows. In Section~\ref{sec:theory}, we introduce the gyrokinetic framework. After specifying the magnetic geometry and field-line-following coordinates, we derive the evolution equation for the zonal fields and obtain expressions for the associated turbulent stresses. In Section~\ref{sec:compute}, we show evidence for the stress competition from local, flux-tube gyrokinetic simulations for different magnetic equilibria. We then draw from simulations an empirical scaling for the ratios of the stresses and demonstrate that the threshold for the non-zonal transition can be predicted from a single, effectively electrostatic, nonlinear simulation. Finally, a brief summary of our work is provided in Section~\ref{sec:conclusion}.

\section{Theory of turbulent stresses}
\label{sec:theory}

\subsection{Gyrokinetics}
We consider turbulent fluctuations that are described by the gyrokinetic--Maxwell system of equations \citep{Catto_1978, Abel_2013} in the absence of strong equilibrium $\bm{E}\times\bm{B}$ rotation of the plasma. This system of equations is obtained by imposing the gyrokinetic ordering on the Vlasov--Fokker--Planck--Maxwell equations and taking advantage of the space-time scale separation between the turbulence and the equilibrium. The associated ordering is
\begin{equation}
\label{eq:gk_ordering}
\frac{\kpar}{\kperp} \sim \frac{|\delta \bm{B}|}{B} \sim \frac{e\phi}{T_s} \sim \frac{\delta f_s}{f_s} \sim \frac{\omega}{\Omega_s} \sim \frac{\rho_s}{L} \equiv \rho_* \ll 1,
\end{equation}
where $\kpar$ and $\kperp$ are the characteristic wavenumbers along and across the equilibrium magnetic field, $\delta \bm{B}$ is the magnetic-field perturbation, $B$ is the strength of the equilibrium magnetic field, $e$ is the elementary charge, $T$ is the temperature, the subscript `$s$' refers to the particle species, $\phi$ is the electrostatic-potential perturbation, $f_s$ is the particle distribution function, with $\delta f_s$ its fluctuating component, $\omega$ is the characteristic frequency of the turbulent fluctuations, $\Omega_s$ and $\rho_s$ are the Larmor frequency and the Larmor radius, respectively, and \mbox{$L \sim L_B \sim L_T \sim L_n$} is a typical scale length of the equilibrium, where $n$ is the density, and $L_{\sigma} = |\grad \ln{\sigma}|^{-1}$ for any equilibrium quantity $\sigma$. In terms of $(\vpar,\,\mu)$ coordinates, where $\vpar$ is the velocity component parallel to the magnetic field and $\mu = m_s\vperp^2/2B$ is the magnetic moment with $\vperp$ the velocity component perpendicular to the magnetic field, the resulting gyrokinetic equation can be written as
\begin{align}
\label{eq:gk_eq}
\frac{\partial h_s}{\partial t} + \vpar \unit{b}\cdot(\grad z)\frac{\partial h_s}{\partial z} + \bm{v}_{Ms} \cdot \gradperp h_s - \frac{\mu}{m_s}(\unit{b}\cdot\grad B) \frac{\partial h_s}{\partial \vpar} + \gyroRs{\vchi}\cdot \gradperp h_s \nonumber \\ + \gyroRs{\vchi}\cdot \grad F_s  =  \frac{Z_s e}{T_s}F_s \frac{\partial \gyroRs{\chi}}{\partial t} + \gyroRs{C\left[h_s\right]},
\end{align}
which describes the time evolution of the non-Boltzmann part of $$\delta f_s = -\frac{Z_se\phi}{T_s}F_s + h_s,$$ where $Z_s$ is the atomic number ($-1$ for electrons). Here $F_s$ is a Maxwellian velocity distribution
\begin{equation}
F_s\left(\bm{R}_s,v\right) = \frac{n_s(\bm{R}_s)}{\left[\pi \vths^2(\bm{R}_s)\right]^{3/2}} \exp{\left[-\frac{v^2}{\vths^2(\bm{R}_s)}\right]},
\end{equation}
where $\bm{R}_s$ is the guiding-centre position, $v$ is the particle speed, and $\vths = \sqrt{2T_s/m_s}$ is the thermal speed with $m_s$ the particle mass. The drift velocities due to the fields' fluctuations and to the equilibrium magnetic field's variation that appear in (\ref{eq:gk_eq}) are given~by
\begin{equation}
\label{eq:vchi_definition}
\gyroRs{\bm{v}_{\chi}} = \frac{c}{B}\unit{b}\times\frac{\partial \gyroRs{\chi}}{\partial \bm{R}_s},
\end{equation}
and
\begin{equation}
\label{eq:vMs_definition}
\bm{v}_{Ms} = \frac{1}{\Omega_s} \left[\left(\vpar^2 + \frac{\vperp^2}{2}\right)\unit{b}\times \grad \ln B + \frac{\beta}{2}\vpar^2 \unit{b}\times \grad \ln p\right],
\end{equation}
respectively, where $c$ is the speed of light, $\unit{b} = \bm{B}/B$, and $\left< \cdots \right>_{\bm{R}_s}$ denotes the gyroaverage at fixed $\bm{R}_s$. The generalised potential in (\ref{eq:vchi_definition}) is
\begin{equation}
\label{eq:chi_definition}
\chi = \phi - \frac{\bm{v}\cdot \bm{A}}{c},
\end{equation}
where $\bm{v}$ is the velocity vector and $\bm{A}$ is the fluctuating vector potential, viz., $\delta \bm{B} = \grad \times \bm{A}$. In (\ref{eq:vMs_definition}), $\beta = 8\pi p/B^2$ is the plasma beta with $p$ the total plasma pressure. In (\ref{eq:gk_eq}), $z$ is the coordinate along the field line and the term $\gyroRs{C\left[h_s\right]}$ is the gyroaveraged collision operator. Unless stated otherwise, the distribution functions $h_s$ and $F_s$ are functions of $\bm{R}_s$, $\vpar$ and $\mu$, while the field quantities $\phi$ and $\bm{A}$ are functions only of the particle position~$\bm{r}$. The gradient operator $\grad$ is defined, for functions of $\bm{R}_s$, as
\begin{equation}
\grad h_s(\bm{R}_s) \equiv \frac{\partial}{\partial \bm{R}_s},    
\end{equation}
and $\grad_{\perp}$ is the gradient operator acting in the plane perpendicular to $\bm{B}$. 

The gyrokinetic equation (\ref{eq:gk_eq}) determines the evolution of the perturbed distribution function $h_s$ given the field equations that determine $\phi$ and $\bm{A}$. The quasi-neutrality constraint gives an equation for $\phi$:
\begin{equation}
\label{eq:QN}
0 = \sum_s Z_s \delta n_s = \sum_s Z_s \left(-\frac{Z_s e \phi}{T_s}n_s + \int \sd^3 \bm{v} \,\left\langle h_s \right\rangle_{\bm{r}}\right),
\end{equation}
where $\delta n_s$ is the density perturbation and $\left<\cdots\right>_{\bm{r}}$ denotes the gyroaverage at fixed $\bm{r}$. The parallel component $A_\parallel$ of the vector potential is calculated from the parallel component of Amp\`{e}re's law:
\begin{equation}
\label{eq:parallel_ampere}
\gradperp^2 \apar = -\frac{4\pi}{c}\sum_s Z_s e \int \sd^3 \bm{v}\,\vpar \gyror{h_s},
\end{equation}
while the perpendicular component $\bm{A}_{\perp}$, related to the field-strength perturbation  $\bpar = \unit{b} \cdot \delta \bm{B} = \unit{b}\cdot \left(\gradperp\times \bm{A}_{\perp}\right)$,
is determined from the perpendicular part of Amp\`{e}re's law:
\begin{equation}
\label{eq:perp_ampere}
\gradperp^2 \bm{A}_{\perp} = - \frac{4\pi}{c} \sum_s Z_s e \int \sd^3 \bm{v}\,\gyror{\bm{v}_{\perp}h_s}.
\end{equation}

Due to the assumed timescale separation between the evolution of the fluctuations and equilibrium quantities, the latter will be held fixed throughout our analytical treatment and in all numerical simulations.

\subsection{Flux-tube coordinates}
\label{sec:flux_coordinates}
We choose to describe the equilibrium magnetic field using the Clebsch representation $\bm{B} = \grad \alpha \times \grad \psi$, in which $\psi$ is the poloidal magnetic flux that labels flux surfaces, defined in such a way that $\psi = 0$ on the magnetic axis. Then $\alpha$ is a function that labels different field lines within a given flux surface and, in axisymmetric configurations, is related to the usual toroidal and straight-field-line poloidal angles $\zeta$ and $\theta_s$, respectively, by $\alpha = \zeta - q\left(\psi\right)\theta_s $, where $q$ is the safety factor. The coordinate along the field line is labelled by $z$, the possible choices of which include, but are not limited to, the toroidal and poloidal angles, and the distance along the field line. We take $z = 0$ to correspond to the outboard midplane of the device. Given that we are interested in working in the local approximation \citep{Beer_1995}, we find it convenient to specialise to field-aligned, flux-tube coordinates $(x,\,y,\,z)$ in which
\begin{equation}
x = \frac{\sd x}{\sd \psi}\left(\psi - \psi_0\right)\quad \mathrm{and} \quad y = \frac{\sd y}{\sd \alpha}\left(\alpha - \alpha_0\right)
\end{equation}
are, respectively, the radial and binormal coordinates. Here, the constants $\psi_0$ and $\alpha_0$ are the values of $\psi$ and $\alpha$ at the centre of the flux tube. The constant derivatives $\sd x/\sd\psi = q/rB_r$ and $\sd y/\sd \alpha = (\sd \psi/\sd r)/B_r$ are chosen so that both $x$ and $y$ have dimensions of length, where $r$ is the half-diameter of the flux surface at the height of the magnetic axis and $B_r$ is a reference magnetic-field strength. In these coordinates, the perpendicular derivative $\gradperp$ of a perturbed quantity $g$ is written as
\begin{equation}
\label{eq:gradperp_definition}
\gradperp g = \grad x\frac{\partial g}{\partial x} + \grad y\frac{\partial g}{\partial y}.
\end{equation}
Note that in axisymmetric devices, equilibrium quantities are independent of the field-line label $\alpha$, and thus of the coordinate $y$. In this work, we define $z$ so that $\bm{B}\cdot \grad z > 0$, so $\grad \alpha \times \grad \psi \cdot \grad z > 0$ or $\grad y \times \grad x \cdot \grad z > 0$ (left-handed coordinates). We are interested in ion-scale turbulence, and the ion Larmor radius $\rho_i = \vthi/\Omega_i$ is typically small compared with the macroscopic length scales of the tokamak, e.g., the device size. We thus adopt the local, flux-tube approximation, in which statistical periodicity of fluctuations in the perpendicular $(x,\,y)$ domain is assumed (provided that the domain size contains at least several correlation lengths of the turbulence).

\subsection{Equation of zonal-flow evolution}
In this section, we derive the evolution equation for the zonal flows, incorporating the interactions with the non-zonal (turbulent) fluctuations. Zonal flows are $\bm{E}\times \bm{B}$ flows that are self-generated by the turbulence and which only vary in $x$. The zonal-flow velocity is
\begin{equation}
v^{\rm{Z}} = -\avg{\frac{c}{B}|\grad x|\frac{\partial \phi}{\partial x}}{yz},
\end{equation}
where the flux-surface average is defined as follows:
\begin{align}
\label{eq:flux_ave_yz}
\avg{\cdots}{yz} = \frac{\int \sd z \, J \avg{\cdots}{y}}{\int \sd z\, J},
\end{align}
where $J = [(\bm{B}\cdot \grad z)(\sd y/d\alpha)(\sd x/\sd \psi)]^{-1}$ is the Jacobian and $\avg{\cdots}{y} = (1/L_y) \int \sd y$ with $L_y$ the domain size in $y$ of the flux tube. Note that we have made use of the local approximation to ignore the slow $(x,\,y)$ variation of $J$. Zonal quantities are then those that have undergone the average \eqref{eq:flux_ave_yz}, after which they become functions purely of $x$.

To obtain the time evolution for $v^{\rm{Z}}$, we first take the time derivative of the quasineutrality constraint (\ref{eq:QN}) and apply the flux-surface average \eqref{eq:flux_ave_yz} to get
\begin{equation}
\label{eq:phi_zonal_evo1}
 \frac{\partial }{\partial t}\sum_s n_s \frac{Z^2_s e}{T_s} \avg{\phi }{yz} = \sum_s Z_s\avg{ \int \sd^3 \bm{v}\,\frac{\partial \gyror{h_s}}{\partial t}}{yz}.
\end{equation}
Substituting the gyrokinetic equation \eqref{eq:gk_eq} into the right-hand side of \eqref{eq:phi_zonal_evo1}, we obtain an evolution equation for the zonal potential (see Appendix~\ref{app:derivation_details}):
\begin{equation}
\label{eq:phi_zonal_evo2}
\frac{\partial }{\partial t}\sum_s \frac{Z^2_s e}{T_s} \avg{\int \sd^3 \bm{v}\, F_s  \left(\phi - \gyror{\gyroRs{\phi}}\right)}{yz}  = \Pi_{\mathrm{lin}} + \Pi_{\phi} + \Pi_{\apar} + \Pi_{\bpar} + \Pi_{\mathrm{coll}},
\end{equation}
where
\begin{align}
\Pi_{\mathrm{lin}}
  &= - \sum_s Z_s \avg{ \int \sd^3\bm{v}\,\left(\bm{v}_{Ms}\cdot \grad x\right)\frac{\partial \gyror{h_s}}{\partial x}}{yz} \nonumber \\
  &\quad - \frac{\partial }{\partial t}\sum_s \frac{Z^2_s e}{cT_s}
     \avg{\int \sd^3 \bm{v}\, F_s \gyror{\gyroRs{\bm{\vperp}\cdot\aperp}}}{yz}, \label{eq:Pi_lin} \\
\Pi_{\phi}
  &= - \sum_s Z_s\avg{\frac{c}{B}\unit{b}\cdot(\grad x\times\grad y)
     \int \sd^3\bm{v}\, \avg{ \left\lbrace \avg{\phi}{\bm{R}_s}, h_s\right\rbrace }{\bm{r}} }{yz}, \label{eq:phi_stress} \\
\Pi_{\apar}
  &= \sum_s Z_s \avg{ \frac{1}{B}\unit{b}\cdot(\grad x\times\grad y)
     \int \sd^3\bm{v}\, \vpar \avg{ \left\lbrace \avg{\apar}{\bm{R}_s}, h_s\right\rbrace }{\bm{r}} }{yz}, \label{eq:apar_stress} \\
\Pi_{\bpar}
  &= \sum_s Z_s\avg{ \frac{1}{B}\unit{b}\cdot(\grad x\times\grad y)
     \int \sd^3\bm{v}\, \avg{\left\lbrace \avg{\bm{v}_{\perp}\cdot\aperp}{\bm{R}_s}, h_s\right\rbrace }{\bm{r}} }{yz}, \label{eq:bpar_stress} \\
\Pi_{\mathrm{coll}}
  &= \sum_s Z_s\avg{ \int \sd^3\bm{v}\, \avg{\gyroRs{C\left[h_s\right]}}{\bm{r}}}{yz} \label{eq:Pi_coll}
\end{align}
are the linear stress, the nonlinear stresses\footnote{Strictly speaking, these quantities have units of particle density per unit time and are therefore not directly comparable to the conventional fluid stresses. Nevertheless, in Appendix~\ref{sec:fluid-limit}, we demonstrate that they can be related to the stresses in the fluid limit.} arising from $\phi$, $\apar$ and $\bpar$, and the collisional damping, respectively. The Poisson bracket is defined as $\llbrace f, g \rrbrace = (\partial f/\partial x)(\partial g/\partial y) - (\partial f/\partial y)(\partial g/\partial x)$. We will refer to the stresses (\ref{eq:phi_stress}) and (\ref{eq:apar_stress}) as kinetic Reynolds and kinetic Maxwell stresses, respectively, to distinguish them from the corresponding fluid versions discussed in Appendix~\ref{sec:fluid-limit}.

To see how (\ref{eq:phi_zonal_evo2}) is connected to the evolution of zonal velocity $v^{\rm{Z}}$, we consider the long-wavelength limit. In the remaining part of this section, we focus on a plasma consisting of two species -- ions with atomic number $Z_i$ and electrons. We do not consider the effects of collisions in this work, so will ignore $\Pi_{\mathrm{coll}}$. Consider the evolution of zonal flows whose radial wavelength is much larger than the ion Larmor radius. These `large-scale' flows often play an important role in the saturation and suppression of turbulence. In this limit, the left-hand side of \eqref{eq:phi_zonal_evo2} can be simplified, and we obtain
\begin{align}
\label{eq:dzonal_dt}
- \frac{\partial}{\partial t}\frac{n_im_i}{e}\frac{\partial^2}{\partial x^2} \avg{\frac{c^2|\grad x|^2}{B^2}\phi}{yz} = \Pi_{\mathrm{lin}} + \Pi_{\phi} + \Pi_{\apar} + \Pi_{\bpar}.
\end{align}
In the limit of large-aspect-ratio, circular flux surfaces, $|\grad x|$ and $B$ are approximately constant in $z$. This enables us to obtain an explicit evolution equation for $(v^{\rm{Z}})^2$. Multiplying both sides of (\ref{eq:dzonal_dt}) by $\avg{e\phi}{yz}$, taking $|\grad x|$ and $B$ to be independent of $z$ and averaging over the $x$ domain yields
\begin{equation}
\label{eq:dzonal_energy_dt}
\frac{\partial}{\partial t}\avg{\frac{n_im_i}{2}(v^{\rm{Z}})^2}{x} = \avg{\avg{e\phi}{yz}\Pi_{\mathrm{lin}}}{x} + \avg{\avg{e\phi}{yz}\Pi_{\phi}}{x} + \avg{\avg{e\phi}{yz}\Pi_{\apar}}{x} + \avg{\avg{e\phi}{yz}\Pi_{\bpar}}{x},
\end{equation}
where $\avg{\cdots}{x} = (1/L_x)\int\sd x$, with $L_x$ the box size in $x$ of the flux tube. This equation describes the time evolution of the box-averaged kinetic-energy density $n_im_i(v^{\rm{Z}})^2/2$ of the zonal flows. The stresses either inject energy into or extract energy from the zonal flows, depending on the sign of the transfer functions appearing on the right-hand side of \eqref{eq:dzonal_energy_dt}. 

In previous studies of electrostatic ITG turbulence, it was found that the kinetic Reynolds stress $\Pi_{\phi}$ acts as a drive for zonal flows \citep{dimits_2007}, although \cite{ivanov_2020} subsequently discovered that, in the fluid limit, the kinetic Reynolds stress $\Pi_{\phi}$ consisted of a fluid Reynolds stress and a diamagnetic stress, of which only the former drove zonal flows while the latter damped them. The kinetic Maxwell stress $\Pi_{\apar}$ as well as some fluid versions of it have been observed to act as zonal-flow-energy sinks \mbox{\citep{Scott_2005, Diamond_2005, Rath_2022, zhang_2026}}. The competition between the kinetic Reynolds and Maxwell stresses, which skews in favour of the kinetic Maxwell stress as beta increases, could lead to net erosion of zonal-flow energy and has thus been hypothesised as the driver behind the non-zonal transition observed in electromagnetic, gyrokinetic simulations \mbox{\citep{Rath_2022, zhang_2026}}. 

\section{Computational study}
\label{sec:compute}
In this section, we present numerical results from gyrokinetic simulations tailored to investigate the effect of the turbulent stresses on the zonal flow in the kinetic regime. We focus in particular on the correlation between the stresses and the non-zonal transition as the plasma beta and safety factor are varied. We use the gyrokinetic code \texttt{stella} \mbox{\citep{Barnes_2019}}, which solves the gyrokinetic-Maxwell system of equations \eqref{eq:gk_eq}, \eqref{eq:QN}, \eqref{eq:parallel_ampere} and \eqref{eq:perp_ampere} in Fourier space. The spatial coordinates are represented using standard pseudo-spectral methods, with Fourier transforms applied in the perpendicular $(x,\,y)$ plane and finite differences used along the field line. Thus, we can expand any fluctuating quantity $g(\bm{r})$ as
\begin{equation}
\label{eq:real_space_rep}
g\left(\bm{r}\right) = \sum_{\bm{k}_{\perp}} g_{\bm{k}_{\perp}}(z) e^{i\bm{k}_{\perp}\cdot \bm{r}} = \sum_{\bm{k}_{\perp}} g_{\bm{k}_{\perp}} \left(z\right) e^{ik_x x + ik_y y},
\end{equation}
which is the representation used in a number of local flux-tube gyrokinetic codes such as \texttt{GS2} \citep{Kotschenreuther_1995}, \texttt{GENE} \citep{Jenko_2000}, \texttt{stella}, \texttt{CGYRO} \citep{candy_2016, candy_2018}, and \texttt{GX} \citep{Mandell_2024}. In \texttt{stella}, as with almost all gyrokinetic codes, the field $\bpar$ is solved for instead of $\bm{A}_{\perp}$. This is because $\gyroRs{\chi}$ can be written in Fourier space as
\begin{equation}
\label{eq:chiR}
\gyroRs{\chi} = \sum_{\bm{k}_{\perp}} e^{i\bm{k}_{\perp}\cdot\bm{R}_s} \Bigg[J_0\left(a_s\right)\left(\phi_{\bm{k}_{\perp}} - \frac{\vpar A_{\parallel,\bm{k}_{\perp}}}{c}\right) + \frac{2\mu}{Z_s e}\frac{J_1\left(a_s \right)}{a_s}B_{\parallel,\bm{k}_{\perp}}\Bigg],
\end{equation}
where $a_s = \kperp \vperp/\Omega_s$ with $\kperp = \left(|\grad x|^2k^2_x + |\grad y|^2k^2_y + 2\grad x \cdot \grad y k_x k_y\right)^{1/2}$, and $J_0$ and $J_1$ are the zeroth and first Bessel functions of the first kind, respectively.  

The electrostatic local version of \texttt{stella} has been successfully benchmarked against \texttt{GS2} for both linear-instability and nonlinear-turbulence calculations \citep{Barnes_2019}. In Appendix \ref{sec:benchmark}, we present additional benchmarking results from comparing three codes -- \texttt{stella}, \texttt{CGYRO} and \texttt{GENE}, now including magnetic fluctuations. 

\begin{table}
\centering
\renewcommand{\arraystretch}{1.2} 
\begin{minipage}[t]{0.48\textwidth}
\centering
\small
\begin{tabular}{|c|c|c|}
\hhline{|---|}
Parameter & CBC & ST40 \\
\hhline{|---|}
$\tilde{r} = r/a$ & $0.5$ & $0.51$ \\
$\tilde{R} = R_0/a$ & $3.0$ & $1.82$ \\
$\sd\tilde{R}/\sd\tilde{r}$ & $0$ & $-0.16$ \\
$\hat{s} = \sd\ln{q}/\sd\ln{r}$ & $0.8$ & $1.18$\\
$\kappa$ & $1.0$ & $1.36$ \\
$\sd\kappa/\sd\tilde{r}$ & $0$ & $0.42$ \\
$\delta$ & $0$ & $0.025$ \\
$\sd\delta/\sd\tilde{r}$ & $0$ & $0.11$\\
$B_r$ & $B_{\zeta}(R_0)$ & $B_{\zeta}(R_0) = 2.41\mathrm{T}$ \\
$n_i/n_r$ & $1$ & $1$ \\
$n_e/n_r$ & $1$ & $1$\\
$T_i/T_r$ & $1$ & $1$\\
$T_e/T_r$ & $1$ & $0.75$\\
$Z_i$ & $1$ & $1$\\
$m_i/m_r$ & $1$ & $1$\\
$m_e/m_r$ & $0.00027$ & $0.00054$\\
$-\sd\ln{n_i}/\sd\tilde{r}$ & $0.733$ & $1.06$ \\
$-\sd\ln{n_e}/\sd\tilde{r}$ & $0.733$ & $1.06$\\
$-\sd\ln{T_i}/\sd\tilde{r}$ & $2.3$ & $2.09$\\
$-\sd\ln{T_e}/\sd\tilde{r}$ & $2.3$ & $1.54$ \\
\hhline{|---|}
\end{tabular}
\caption{Equilibrium parameters for the CBC and the ST40 pulse 314. Here $a$ is the half-diameter of the last-closed flux surface, $\hat{s}$ is the magnetic shear, $B_{\zeta}$ is the toroidal component of the magnetic field, and $n_r$, $T_r$, and $m_r$ are reference density, temperature, and mass, respectively.}
\label{tab:norms}
\end{minipage}%
\hfill
\begin{minipage}[t]{0.48\textwidth}
\centering
\small
\setlength{\tabcolsep}{2pt}
\renewcommand{\arraystretch}{1.2}
\begin{tabular}{|c|c|c|c|c|}
\hhline{|-----|}
Resolution & $L_y/\rho_r$ & $L_x/\rho_r$ & $k_{x,\mathrm{max}}\rho_r$ & $k_{y,\mathrm{max}}\rho_r$ \\
\hhline{|-----|}
\textit{R1} & $94$ & $94$ & $2.8$ & $1.4$ \\
\textit{R2} & $141$ & $141$ & $2.8$ & $1.4$\\
\textit{R3} & $188$ & $188$ & $2.8$ & $1.4$\\
\textit{R4} & $188$ & $188$ & $2.8$ & $1.4$\\
\textit{R5} & $94$ & $94$ & $5.7$ & $2.8$\\
\textit{ST1} & $118$ & $126$ & $3.3$ & $1.6$\\
\textit{ST2} & $89$ & $94$ & $3.0$ & $1.4$\\
\hhline{|-----|}
\end{tabular}
\caption{Resolution of our simulations. The subscript `max' stands for the maximum of the corresponding quantity; $\rho_r = m_rv_r c/eB_r$ is the reference Larmor radius with $v_r = \sqrt{2T_r/m_r}$. Here \textit{R1-5} are the resolutions used for the CBC simulations and \textit{ST1} and  \textit{ST2} are those used for the ST40 cases. We used $N_{\mu} = 16$ for \textit{R1-5} and $N_{\mu} = 12$ for \textit{ST1} and \textit{ST2}. For all the simulations listed here, $v_{\parallel,\mathrm{max}}/\vths = 3$, $v_{\perp,\mathrm{max}}/\vths = 3$,  $N_z = 32$ and $N_{\vpar} = 48$.}
\label{tab:reso}
\end{minipage}
\end{table}

\begin{figure}
    \centering
    \hspace*{-3cm}\text{CBC}\\
    \includegraphics[scale=0.6]{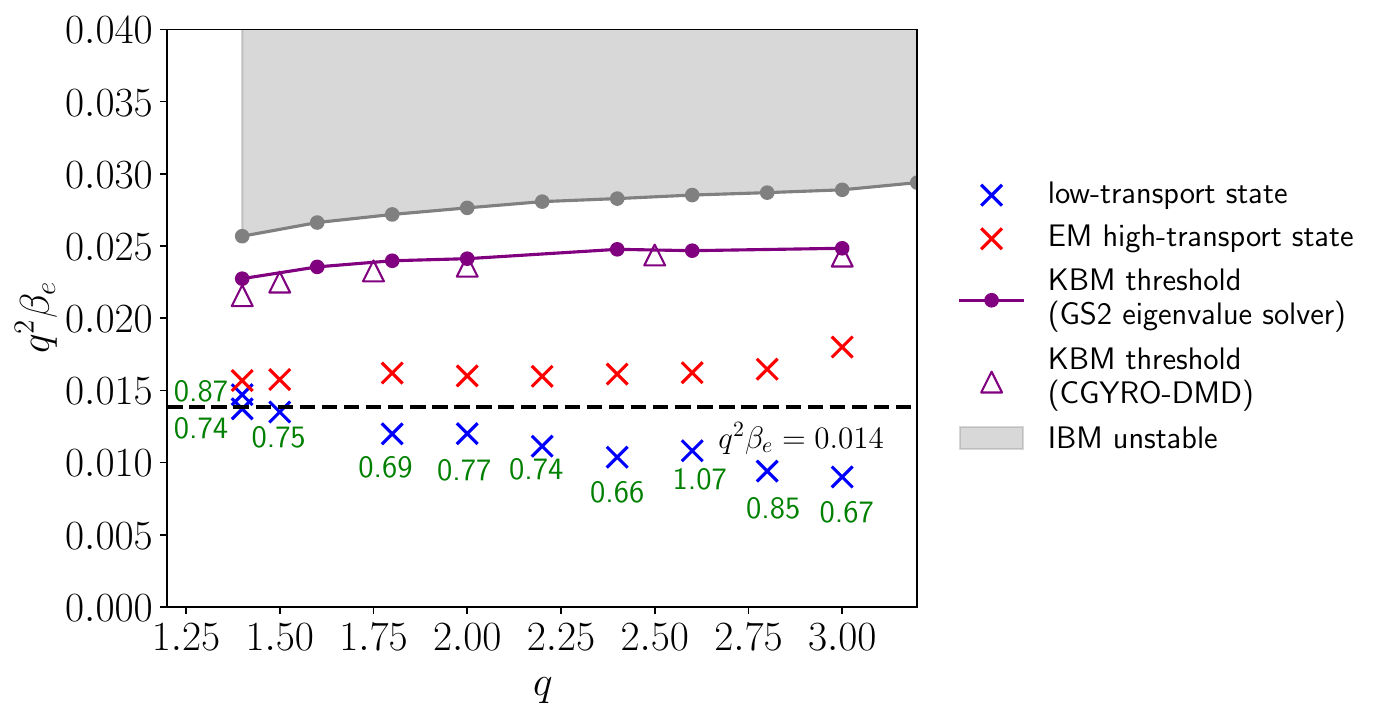}\\[1em]
    \hspace*{-3cm}\text{ST40}\\
    \includegraphics[scale=0.6]{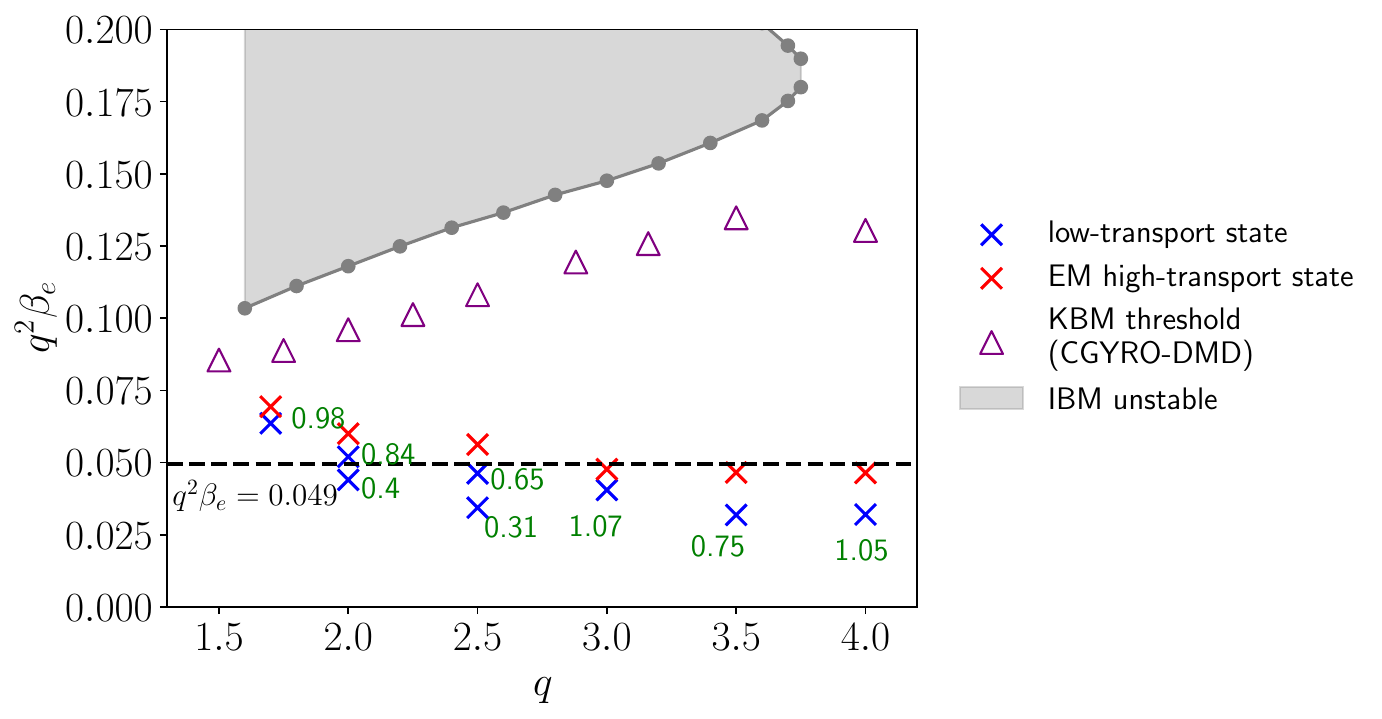}
    \caption[]{The non-zonal-transition boundary and the linear-stability thresholds of KBMs and IBMs in the $(q,\,\beta_e)$ parameter space for the CBC (top) and ST40 (bottom) equilibria. The blue crosses represent cases with strong zonal flows and modest (saturated) heat flux, and the red crosses represent cases with weak zonal flow and large (divergent) heat fluxes. The corresponding values of the energy transfer ratio $r_{\mathrm{MR}}$ between Maxwell and Reynolds stresses, defined by (\ref{eq:ratio_def}), for saturated cases are given in green. The dashed black lines indicate empirical estimates of the non-zonal-transition boundary. For each value of $q$, the transition is bracketed by the highest-$\beta_e$ low-transport simulation and the lowest-$\beta_e$ high-transport simulation. The horizontal line is chosen to provide the best overall representation of these bracketing intervals across the full range of $q$, giving $q^2\beta_e \approx 0.014$ for CBC and $q^2\beta_e \approx 0.049$ for ST40. The purple triangles represent the linear-stability thresholds of KBMs obtained from \texttt{CGYRO-DMD}~\citep{Dudkovskaia_2025}, and the grey areas are IBM-unstable regions. In the top panel, the purple line is the KBM linear-stability threshold obtained from the \texttt{GS2} eigenvalue solver.}
    \label{fig:q_beta_plots}
\end{figure}

We perform our calculations for two different magnetic equilibria. The first equilibrium is the Cyclone-Base Case (CBC), which has circular, concentric flux surfaces and is widely used for benchmark studies (e.g., \citeauthor{dimits_2000} \citeyear{dimits_2000}). The second equilibrium is taken from the ST40 \citep{McNamara_2024, otto_2026} pulse number 314, ASTRA simulation Run 102 at $\mathrm{time}=450\,\mathrm{ms}$ (numerical pulse). These are both parametrised in terms of the Miller local parameters \citep{miller_1998, stacey_2008}. In this approach, the flux surface of interest is described by the cylindrical coordinates $(R, Z)$, with $ R(r, \vartheta) = R_0(r) + r\,\mathrm{cos}\left[\vartheta + \mathrm{sin}\,\vartheta\,\mathrm{arcsin}\,\delta(r)\right]$ and $Z(r, \vartheta) = \kappa(r)r\,\mathrm{sin}\,\vartheta$, where $\vartheta$ is the poloidal-angle-like coordinate and $R_0$ is the major radius. The parameters $\delta$ and $\kappa$ control the triangularity and elongation of the given plasma surface, respectively. The Miller parameters as well as other equilibrium parameters for both equilibria are provided in Table~\ref{tab:norms}.

The spatial coordinates used in \texttt{stella} are $(k_x,\,k_y,\,z)$, while the velocity-space coordinates are $(\vpar,\,\mu)$. Therefore, the resolution inputs for the simulations are the box sizes $L_x$ and $L_y$ in $x$ and $y$, respectively (the $z$-domain is limited to [$-\pi$, $\pi$] in our simulations), the maximum parallel velocity component $v_{\parallel,\mathrm{max}}/\vths$, the maximum perpendicular velocity component $v_{\perp,\mathrm{max}}/\vths$, the number $N_{k_x}$ of Fourier modes in $x$, the number $N_{k_y}$ of Fourier modes with positive $k_y$, the number $N_z$ of $z$-grid points, the number $N_{\vpar}$ of $\vpar$-grid points, and the number $N_{\mu}$ of $\mu$-grid points. The values that we have used for these parameters are given in Table~\ref{tab:reso}.

For all simulations reported here, physical collisions are neglected. Instead, hyper-viscosity of the form
\begin{equation}
\label{eq:hyper}
\left(\frac{\partial h_{s,\bm{\kperp}}}{\partial t}\right)_{\mathrm{hyper}} = - D_{\mathrm{hyper}}\left(\frac{\kperp}{k_{\perp,\mathrm{max}}}\right)^4 h_{s,\bm{\kperp}}
\end{equation} 
is employed, where $D_{\mathrm{hyper}}$ is the hyper-viscosity coefficient.

\subsection{Non-zonal transitions in CBC and ST40}
\label{sec:nonzonal-cbc-st40}
 
The data from a $(q,\,\beta_e)$ parameter scan are shown in Fig.~\ref{fig:q_beta_plots}, where the vertical axis is $q^2\beta_e$ and the horizontal axis is $q$, for both the CBC (top panel) and ST40 (bottom panel) equilibria. This representation highlights the various important boundaries in parameter space. The ratio $\beta_i/\beta_e$ is kept fixed as $\beta_e$ is varied, where $\beta_{i,e}=8\pi n_{i,e}T_{i,e}/B_r^2$ and $B_r$ is the reference magnetic field, defined as the toroidal component of the magnetic field at $R = R_0$ on the flux surface. For the CBC runs, we set $\beta^{\prime} \equiv (4\pi/B^2_r) dp/d\tilde{r} = 0$ (here $\tilde{r} = r/a$ with $a$ the half-diameter of the last-closed flux surface), $D_{\mathrm{hyper}} = 0.2$ and neglect $\bpar$. We used resolution \textit{R1} (see Table~\ref{tab:reso}) for $q=1.4,\,1.5,$ and $1.8$, resolution \textit{R2} for $q=2.0,\,2.2,$ and $2.4$, and resolution \textit{R3} for $q=2.6,\,2.8,$ and $3.0$. This corresponds to an increase in the $x$ and $y$ box sizes with $q$, needed to account for the shift in outer scale with $q$~\citep{Barnes_2011, Adkins_2022, Adkins_2023, nies_2024, Ivanov_2025,adkins2026}. For the ST40 runs, we retain non-zero $\beta^{\prime}$ (consistent with other parameters) and $\bpar$, and let $D_{\mathrm{hyper}} = 2.0$. A single resolution (\textit{ST1} in Table~\ref{tab:reso}) was used for all the ST40 runs shown in Fig.~\ref{fig:q_beta_plots}, which appeared to be sufficient to resolve the outer scales.

The non-zonal-transition boundary lies within the spacing between the blue and red crosses, which represent low- and high-transport states, respectively. For the CBC cases, the transition boundary occurs at an approximately constant value of $q^2\beta_e$, 
\[
q^2\beta_e = C_{\mathrm{nl}},
\]
where $C_{\mathrm{nl}} \approx 0.014$. To probe the role played by linear instabilities, we also performed a series of linear simulations to calculate the stability thresholds of KBMs and ideal ballooning modes (IBMs). The KBM linear-stability threshold, shown as the purple line in the top panel (obtained using the \texttt{GS2} eigenvalue solver) and the purple triangles in both panels (obtained using \texttt{CGYRO-DMD}), also occurs at an approximately constant value of $q^2\beta_e$,
\[
q^2\beta_e = C_{\mathrm{KBM}},
\]
where $C_{\mathrm{KBM}} \approx 0.024$. Similarly, the lower IBM linear-stability threshold, shown as the grey line, satisfies
\[
q^2\beta_e = C_{\mathrm{IBM}},
\]
with $C_{\mathrm{IBM}} \approx 0.028$. Thus, although the linear-stability thresholds also exhibit an approximate $q^2\beta_e$ scaling, they satisfy
\[
C_{\mathrm{nl}} < C_{\mathrm{KBM}} < C_{\mathrm{IBM}},
\]
indicating that neither the subdominant KBMs nor the IBMs are responsible for triggering the non-zonal transition.

For the ST40 cases, the same qualitative ordering, $C_{\mathrm{nl}} < C_{\mathrm{KBM}} < C_{\mathrm{IBM}}$, is retained, with $C_{\mathrm{nl}} \approx 0.049$, $C_{\mathrm{KBM}} \approx 0.10$, and $C_{\mathrm{IBM}} \approx 0.14$. However, unlike the CBC results, both the linear-stability thresholds and the non-zonal-transition boundary seem to have noticeable deviation from the $q^2\beta_e=\mathrm{constant}$ scaling. Notice also that the KBM linear-stability threshold exhibits a different trend from the non-zonal-transition boundary, providing even stronger evidence that the subdominant KBMs are not responsible for triggering the non-zonal transition. Thus, the linear-stability thresholds provide very limited predictive power for the non-zonal-transition boundary, particularly for the ST40 cases.

\subsection{Turbulent energy transfers}
\label{sec:turbulent energy transfers}
As the system undergoes the non-zonal transition, zonal activity weakens significantly, leading to an EM high-transport state. To explore this, we examine the zonal drives directly by calculating the turbulent stresses. In particular, we focus on the effect of turbulent stresses on the large-scale zonal flows. We first write the evolution equation for the large-scale, zonal potential \eqref{eq:dzonal_dt} in Fourier space:
\begin{equation}
\label{eq:dzonalkx_dt}
\frac{\partial}{\partial t}\frac{n_im_i}{e} \frac{c^2|\grad x|^2k^2_x}{B^2}\avg{\phi}{yz,k_x} = \Pi_{\mathrm{lin},k_x} + \Pi_{\phi,k_x} + \Pi_{\apar,k_x} + \Pi_{\bpar,k_x},
\end{equation}
where the subscript `$k_x$' refers to the $k_x$ component of the quantity, and we have again assumed that $\grad x$ and $B$ are approximately constant along $z$\footnote{This approximation does not apply to the ST40 cases due to their more complicated geometry. The purpose of this exercise is to illustrate the physical meaning of the energy transfers.}. To obtain an equation for the scale-by-scale energy transfer, we multiply both sides of \eqref{eq:dzonalkx_dt} by $\avg{e\phi}{yz,k_x}^*$ and add to the resulting equation its own complex conjugate. The resulting evolution equation for the component of the zonal-flow energy at wavenumber $k_x$ is
\begin{equation}
\label{eq:transfer}
\frac{\partial}{\partial t} \frac{n_im_i}{2} \left|v^{\rm{Z}}_{k_x}\right|^2 = T_{\mathrm{lin},k_x} + T_{\phi,k_x} + T_{\apar,k_x} + T_{\bpar,k_x},
\end{equation}
where we have defined the energy transfers from the turbulence to the zonal flows $T_{\mathrm{lin},k_x} = \mathrm{Re}(\avg{e\phi}{yz,k_x}^*\Pi_{\mathrm{lin},k_x})$, $T_{\phi,k_x} = \mathrm{Re}(\avg{e\phi}{yz,k_x}^*\Pi_{\phi,k_x})$, $T_{\apar,k_x} = \mathrm{Re}(\avg{e\phi}{yz,k_x}^*\Pi_{\apar,k_x})$, and $T_{\bpar,k_x} = \mathrm{Re}(\avg{e\phi}{yz,k_x}^*\Pi_{\bpar,k_x})$. Notice that the left-hand side of (\ref{eq:transfer}) is a time derivative of a positive-definite quantity; the signs of the transfer functions on the right-hand side indicate whether the associated stress injects energy into or extracts it from the zonal mode at a scale $k_x$.

\begin{figure}
	\centering
	\begin{center}
		\begin{tabular}{ cc } 
			\includegraphics[scale=1.0]{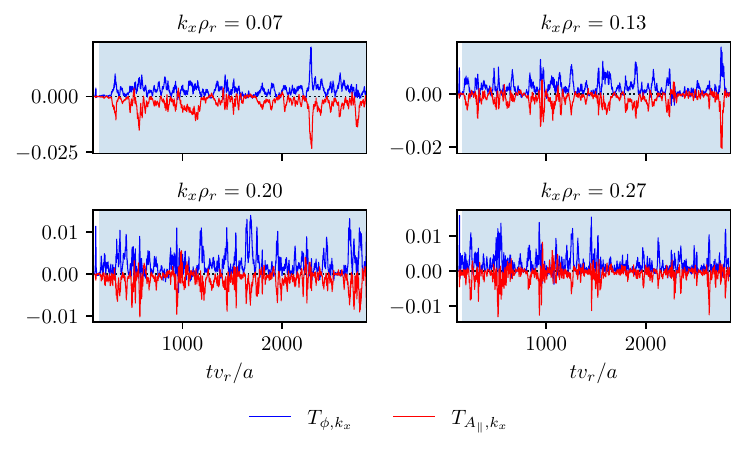}
		\end{tabular}
	\end{center}
	\caption[]{Time traces of the energy transfers due to the Reynolds ($T_{\phi,k_x}$) and Maxwell ($T_{\apar,k_x}$) stresses, normalised by $n_rT_rv_r\rho^2_r/a^3$, for the CBC run with $q=1.4,\,\beta_e = 0.007$ (near the non-zonal-transition boundary), and resolution \textit{R1}. The red and blue curves show the transfers $T_{\apar,k_x}$ and $T_{\phi,k_x}$, respectively. The four panels show the contributions to the transfers from the four lowest $k_x$. The time averages are performed over the time windows indicated by the light-blue shading.}
	\label{fig:transfer_time_traces}
\end{figure}

\begin{figure}
    \centering
    \begin{tabular}{cc}
        \includegraphics[scale=0.4]{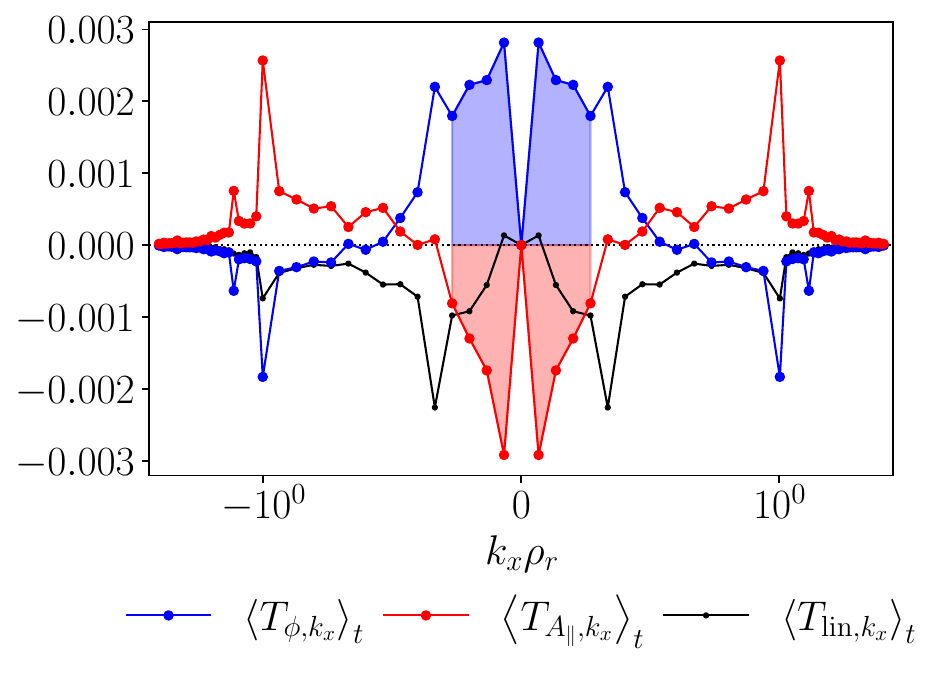} &
        \raisebox{-4mm}{%
            \includegraphics[scale=0.4]{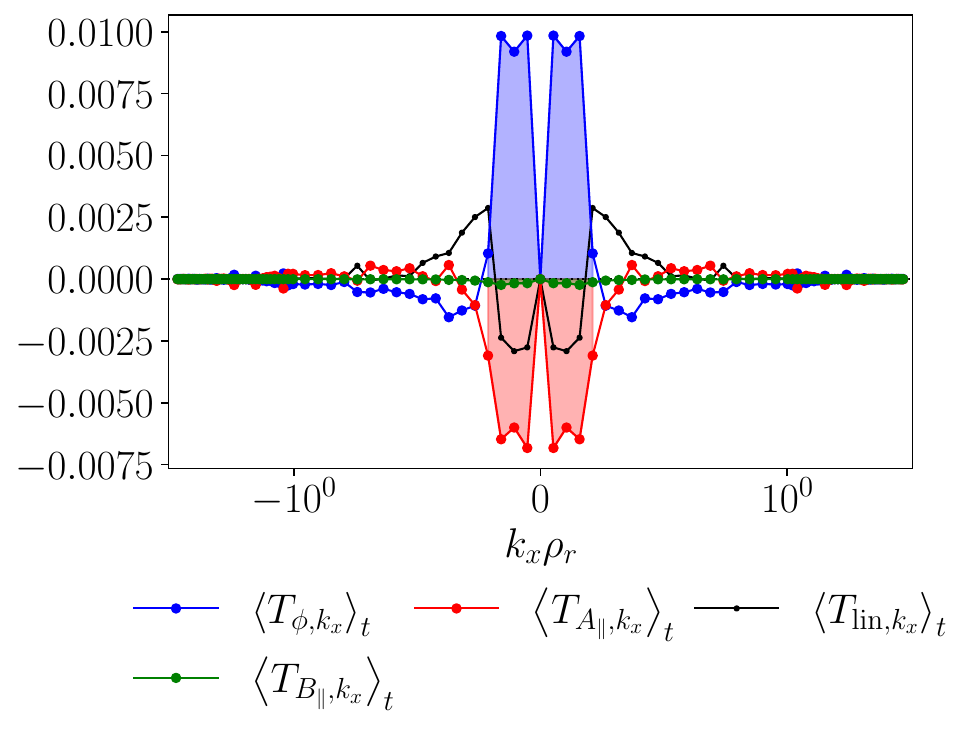}}
    \end{tabular}
    \caption[]{Time-averaged energy-transfer spectra normalised by $n_rT_rv_r\rho_r^2/a^3$. The blue and red regions show the large-scale transfers due to $T_{\phi,k_x}$ and $T_{\apar,k_x}$, respectively. \textbf{Left panel}: CBC with $q=1.4$,\,$\beta_e=0.007$, and resolution \textit{R1}. \textbf{Right panel}: ST40 case with $q=3.5$,\,$\beta_e=0.0026$, and resolution \textit{ST1}.}
    \label{fig:transfer_spec_1}
\end{figure}

Fig.~\ref{fig:transfer_time_traces} shows the time traces of $T_{\phi,k_x}$ (blue curves) and $T_{\apar,k_x}$ (red curves) for the CBC with $q=1.4,\,\beta_e = 0.007$, and resolution \textit{R1}. The four panels show the contributions to the transfers from the four lowest $k_x$. A saturated state where the transfers become quasi-steady is clearly identifiable. Note that the Reynolds and Maxwell transfers are comparable for this case, which is near the non-zonal-transition boundary, suggesting a competition between them. Once we obtain the time evolution of $T_{\sigma,k_x}$ where $\sigma$ stands for one of the stresses, we can calculate the time-averaged transfer spectrum in the $k_x$-space (see Fig.~\ref{fig:transfer_spec_1}), revealing how different scales contribute to the zonal flow dynamics. In the left panel of Fig.~\ref{fig:transfer_spec_1}, we show data from the CBC with $q = 1.4,\,\beta_e = 0.007$, and resolution \textit{R1}. While the Reynolds transfer $T_{\phi,k_x}$ remains positive at small $k_x$, indicating a zonal drive at large scales, it becomes negative at higher $k_x$, with sharp structures near $k_x \sim 1$. These complicated structures are sensitive to numerical resolution and disappear when the resolution is increased (further justification of focusing on large-scale transfers is given in Appendix~\ref{sec:large_scale}). In the right panel, we show data from the ST40 simulation with $q=3.5,\,\beta_e = 0.0026$, and resolution \textit{ST1}. There, the complicated high-$k_x$ structures are absent and the large-scale transfers are clearly the dominant contribution, with the signs of Reynolds (positive) and Maxwell (negative) transfers matching the expectations.

To study quantitatively how the Maxwell-Reynolds competition correlates with the non-zonal transition, we define the time-averaged, large-scale transfer $T_{\sigma}$ using the low-pass wavenumber filter:
\begin{equation}
\label{eq:large_scale_transfer}
\avg{T_{\sigma}}{t} = \sum_{|k_x| \leq k_{\mathrm{cut}}} \avg{T_{\sigma,k_x}}{t},
\end{equation}
where $k_{\mathrm{cut}}$ is the $k_x$ at which either of the transfers changes sign, i.e., the Reynolds (Maxwell) transfer switches from positive (negative) to negative (positive). To summarise, for a given simulation, one finds $\avg{T_{\sigma}}{t}$ by the following steps: (i) calculate transfer functions $T_{\sigma,k_x}$; (ii) calculate the time-averaged\footnote{In certain cases, the long-term evolution of the turbulence may complicate the time average. See Appendix~\ref{sec:long_evolution} for more details.} transfer functions $\avg{T_{\sigma,k_x}}{t}$ over the saturation period; (iii) find $k_{\mathrm{cut}}$; and (iv) evaluate the sum $\sum_{|k_x| \leq k_{\mathrm{cut}}} \avg{T_{\sigma,k_x}}{t}$ to find the large-scale transfer $\avg{T_{\sigma}}{t}$.

As we anticipate a competition between Maxwell and Reynolds stresses, we consider the Maxwell-to-Reynolds energy-transfer ratio 
\begin{equation}
\label{eq:ratio_def}
r_{\mathrm{MR}}\equiv \left|\frac{ \avg{T_{\apar}}{t}}{\avg{T_{\phi}}{t}}\right|
\end{equation}
(also shown as the ratio of red area to blue area in Fig.~\ref{fig:transfer_spec_1}). The value of $r_{\mathrm{MR}}$ is zero when $\beta_e = 0$ and increases with $\beta_e$. As $r_{\mathrm{MR}}$ reaches unity or exceeds it, the Maxwell transfer cancels the Reynolds transfer and the zonal flows cease to be sustained. The transfer ratio for each saturated simulation is labelled in Fig.~\ref{fig:q_beta_plots} by green numbers. As one can see, the transfer ratio approaches unity near the non-zonal-transition boundary in both CBC and ST40, as expected.

\subsection{Scaling of the turbulent energy-transfer ratio with $\beta_e$}
Let us now investigate the scaling of $r_{\mathrm{MR}}$ with $q$ and $\beta_e$ in electromagnetic ITG turbulence. We will show that this scaling can be used to predict the non-zonal-transition boundary in the $(q,\beta_e)$ plane from a single nonlinear simulation performed in the approximately electrostatic limit. Such a simulation is computationally much less expensive than one carried out near the transition boundary, where the rapid electron motion along perturbed magnetic field lines imposes a stringent time-step constraint for numerical stability.

\subsubsection{Analytical estimate}
\label{sec:analytical_estimate}
The ratio $r_{\mathrm{MR}}$ can be estimated from the expressions for the kinetic Reynolds (\ref{eq:phi_stress}) and Maxwell (\ref{eq:apar_stress}) stresses. Assuming that the characteristic spatial scales of the two stresses are comparable, we obtain
\begin{equation}
\label{eq:ratio_estimate1}
r_{\mathrm{MR}} \sim \frac{\Pi_{\apar}}{\Pi_{\phi}} \sim  \frac{\apar \sum_s Z_s\int \sd^3\bm{v}\,\vpar \gyror{h_s}}{c\phi \sum_s Z_s\int \sd^3\bm{v}\, \gyror{h_s}} \sim \frac{T_e}{ \rho^2_i n_e e^2}\frac{\apar^2}{\phi^2} \sim \frac{B^2}{ m_i n_e c^2}\frac{\apar^2}{\phi^2},   
\end{equation}
where we have used quasineutrality (\ref{eq:QN}), parallel Amp\`{e}re's law (\ref{eq:parallel_ampere}), and $\kperp \rho_i \sim T_i/T_e\sim 1$ (independent of $\beta_e$). Thus, estimating $r_{\mathrm{MR}}$ reduces to estimating the ratio~$\apar/\phi$. 

To do this, we first consider the evolution equation for the electron density, obtained by taking the density moment $(1/n_e)\int \sd^3\bm{v}\,\gyror{\cdots}$ of the gyrokinetic equation (\ref{eq:gk_eq}) for electrons:
\begin{equation}
\label{eq:electron_density}
\frac{\partial}{\partial t}\frac{\delta n_e}{n_e} + (\bm{B}\cdot\grad z)\frac{\partial}{\partial z}\frac{u_{\parallel,e}}{B} + \cdots = 0,
\end{equation}
where $\delta n_e$ is the electron-density fluctuation, $u_{\parallel,e} = (1/n_e)\int \sd^3\bm{v}\,\vpar\gyror{h_e}$ is the electron parallel flow, and `$\cdots$' denotes the remaining terms in the electron-density equation that are not relevant to the present argument. In the electrostatic limit, the non-zonal part of the electron-density fluctuation is expected to follow the Boltzmann response, \mbox{$\delta n_e/n_e = e\phi/T_e$}. This means that, with the evolution of $\phi$ determined via the quasineutrality condition (\ref{eq:QN}), the electron density must evolve consistently with that. This consistency is maintained by the parallel electron flow $u_{\parallel,e}$, which adjusts the density evolution through the parallel-compressibility term. Consequently, for small values of $\beta_e$, we still expect \mbox{$\delta n_e/n_e \sim e\phi/T_e$} and the two retained terms in (\ref{eq:electron_density}) to balance, which yields
\begin{equation}
\label{eq:ratio_upare_phi}
\frac{u_{\parallel,e}/\vthi}{e\phi/T_e} \sim \frac{\omega}{\kpar \vthi} \sim q\sqrt{\frac{R}{L_{T_i}}},
\end{equation}
where we have used $\omega \sim k_y\rho_i\vthi/\sqrt{R L_{T_i}}$ (typical growth rate of curvature ITG instabilities) with $L_{T_i}$ the ion-temperature-gradient scale length, $k_y\rho_i \sim 1$, and $\kpar \sim 1/qR$. 

The parallel Amp\`ere's law gives
\begin{equation}
\label{eq:parallel_amp_apar_phi}
\frac{\apar}{\phi} \sim \frac{e}{c\phi\kperp^2}\int \sd^3\bm{v} \,\vpar\gyror{h_e} \sim \frac{m_i n_e \vthi c}{ B^2}\frac{u_{\parallel,e}/\vthi}{e\phi/T_e},
\end{equation}
or, in terms of normalised amplitudes and using (\ref{eq:ratio_upare_phi}),
\begin{equation}
\label{eq:ratio_aparphi_normed}
\frac{\apar/\rho_i B}{e\phi/T_e} \sim \frac{\apar\vthi}{\phi c} \sim q \beta_e \frac{R}{L_{T_i}}.
\end{equation}
Combining (\ref{eq:parallel_amp_apar_phi}) with (\ref{eq:ratio_estimate1}) and (\ref{eq:ratio_upare_phi}), we get
\begin{equation}
\label{eq:transfer_ratio_scaling_estimate}
r_{\mathrm{MR}} \sim q^2\beta_e \frac{R}{L_{T_i}}.
\end{equation}
This estimate predicts that, for fixed $R/L_{T_i}$, the transfer ratio
$r_{\mathrm{MR}}$ scales as $q^2\beta_e$, consistent with the predictions of a
fluid model for electromagnetic ITG turbulence~\citep{zhang_2026} in a $Z$-pinch, where we found $r_{\mathrm{MR}} \propto (L^2_{\parallel}/L^2_B)\beta_e$, with $L_{\parallel}$ the box size along the magnetic field (in a tokamak, $L_{\parallel} \sim qR$) and $L_B \sim R$ the scale length of the magnetic-field-strength gradient. In the following section, we test this scaling against our gyrokinetic simulations.

\subsubsection{Numerical validation}
\label{sec:numerical_validation}
\begin{figure}
    \centering
    \includegraphics[scale=0.6]{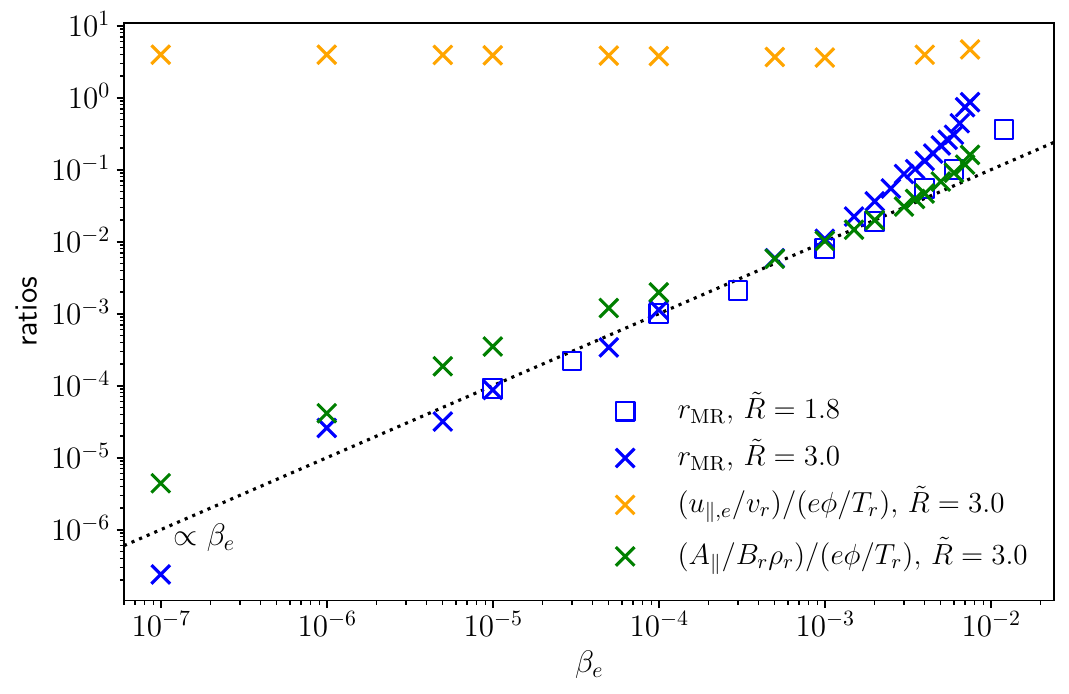}
    \caption[]
    {Scaling with $\beta_e$ of the energy-transfer ratio $r_{\mathrm{MR}}$, defined in (\ref{eq:ratio_def}), as well as the ratios between the normalised amplitudes of the fields, for the CBC simulations with $q=1.4$ and resolution \textit{R1}. The scans are carried out for $\tilde{R} = 1.8$ (blue squares) and $\tilde{R} = 3.0$ (blue crosses). For the simulations with $\tilde{R} = 3.0$, the ratios (\ref{eq:ratio_upare_phi}) (yellow crosses) and (\ref{eq:ratio_aparphi_normed}) (green crosses) between the non-zonal components of the fields are also plotted. They are obtained by summing the squared modulus of each field (e.g., $|\phi_{\bm{k}_{\perp}}|^2$) over $k_x$, $k_y$, and $z$, and then forming ratios. The black dotted line is the theoretical scaling (\ref{eq:transfer_ratio_scaling_estimate}).}
    \label{fig:transfer_ratio_beta}
\end{figure}

In Fig.~\ref{fig:transfer_ratio_beta}, we compare the analytical estimates of
Section~\ref{sec:analytical_estimate} with nonlinear simulations of the CBC. The transfer ratio $r_{\mathrm{MR}}$ is shown for two aspect ratios,
$\tilde{R}=1.8$ and $\tilde{R}=3.0$ (corresponds to the top panel of Fig.~\ref{fig:q_beta_plots}). We also show the normalised ratios $(u_{\parallel,e}/v_{\mathrm{th}i})/(e\phi/T_e)$ and
$(A_{\parallel}/B\rho_i)/(e\phi/T_e)$ for the non-zonal components ($k_y \neq 0$, since they contribute to the stresses) of the fields obtained from simulations with $\tilde{R}=3.0$. The other equilibrium parameters are the same as those in Table~\ref{tab:norms} and the resolution used is \textit{R1}. The black dotted line indicates
the scaling $r_{\mathrm{MR}}\propto\beta_e$.

At sufficiently low $\beta_e$, the simulations agree well with the
analytical estimates. The transfer ratio scales approximately as
$r_{\mathrm{MR}}\propto\beta_e$, while
$(u_{\parallel,e}/v_{\mathrm{th}i})/(e\phi/T_e)$ remains nearly constant
and $(A_{\parallel}/B\rho_i)/(e\phi/T_e)$ also scales linearly with
$\beta_e$, consistent with (\ref{eq:ratio_upare_phi}) and
(\ref{eq:ratio_aparphi_normed}). 

\begin{figure}
    \centering
    \includegraphics[scale=0.6]{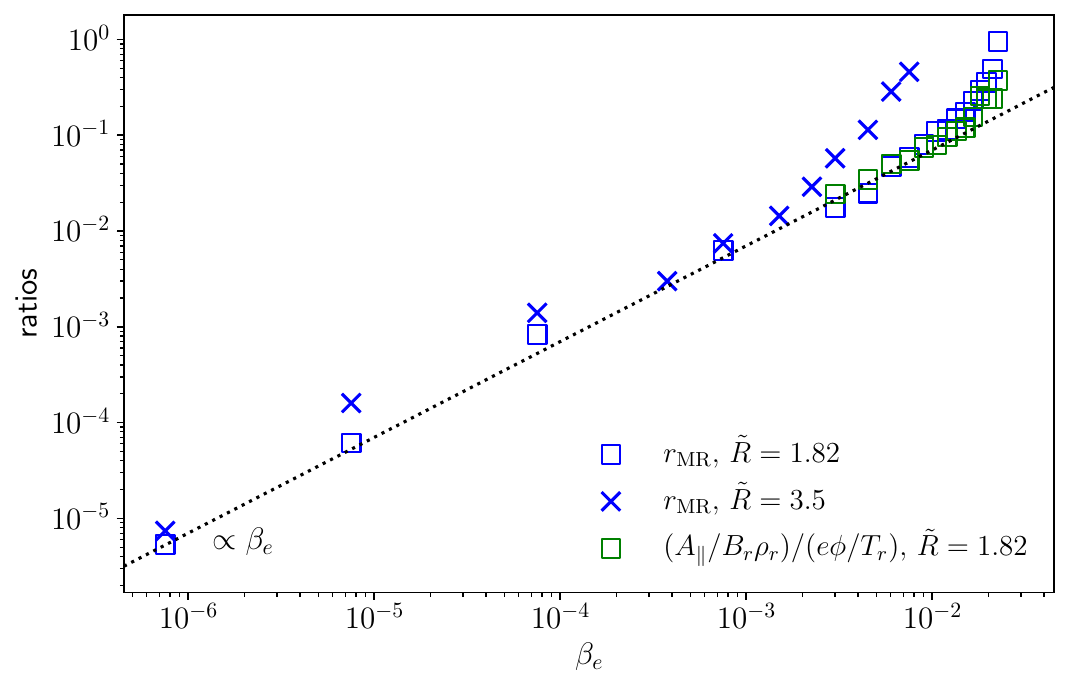}
    \caption[]
    {Scaling with $\beta_e$ of the energy-transfer ratio $r_{\mathrm{MR}}$, defined in (\ref{eq:ratio_def}), as well as the ratios between the normalised amplitudes of the fields, for the ST40 simulations with $q=1.7$ and resolution \textit{ST2}. The scans are carried out for $\tilde{R} = 1.82$ (blue squares) and $\tilde{R} = 3.5$ (blue crosses). The black dotted line is the scaling (\ref{eq:transfer_ratio_scaling_estimate}).}
    \label{fig:transfer_ratio_beta_st40}
\end{figure}

As $\beta_e$ increases beyond a certain critical value, which we denote by
$\beta_{e,\mathrm{sb}}$ (scaling breakdown), $r_{\mathrm{MR}}$
begins to deviate systematically from the predicted linear scaling,
increasing more rapidly than $\beta_e$. This departure occurs while $(u_{\parallel,e}/v_{\mathrm{th}i})/(e\phi/T_e)$
remains relatively constant and
$(A_{\parallel}/B\rho_i)/(e\phi/T_e)$ exhibits a stronger scaling that
closely follows that of $r_{\mathrm{MR}}$, suggesting that the breakdown
of the scaling $r_{\mathrm{MR}}\propto\beta_e$ arises mainly from the failure of
the estimate (\ref{eq:parallel_amp_apar_phi}).

For the ST40 equilibrium, the situation is familiar, Fig.~\ref{fig:transfer_ratio_beta_st40} shows
the dependence of $r_{\mathrm{MR}}$ on $\beta_e$ for two values of the
aspect ratio, $\tilde{R}=1.82$ (corresponds to the bottom panel of Fig.~\ref{fig:q_beta_plots}) and $\tilde{R}=3.5$. The other equilibrium parameters are the same as those in Table~\ref{tab:norms} and the resolution used is \textit{ST2}. As in the CBC, the
analytical prediction $r_{\mathrm{MR}}\propto\beta_e$ is recovered at
sufficiently low $\beta_e$, but as $\beta_e$ increases, $r_{\mathrm{MR}}$ departs systematically from the linear scaling and grows more rapidly. The ratio (\ref{eq:ratio_aparphi_normed}) obtained from the simulations with $\tilde{R} = 1.82$ closely tracks the behaviour of $r_{\mathrm{MR}}$.

A common trend for both the CBC and ST40 simulations is that the value $\beta_{e,\mathrm{sb}}$ where the scaling (\ref{eq:transfer_ratio_scaling_estimate}) breaks down is smaller at larger aspect ratios. In other
words, more compact equilibria retain the asymptotic
$r_{\mathrm{MR}}\propto\beta_e$ scaling over a wider range of $\beta_e$. One possible explanation is that the analytical estimate presented in Sec.~\ref{sec:analytical_estimate} implicitly assumes the system to remain sufficiently close to the electrostatic limit. As the aspect ratio $R$ increases, however, the system becomes increasingly electromagnetic owing to the longer parallel connection length. This can be quantified by comparing the characteristic ITG growth rate with the Alfv\'en transit frequency:
\[
\frac{\vthi/\sqrt{RL_{T_i}}}{v_A/qR}
\propto
q\sqrt{\beta_i\frac{R}{L_{T_i}}},
\]
where $v_A$ is the Alfv\'en speed. Notice that the same combination of parameters also appears in (\ref{eq:transfer_ratio_scaling_estimate}), assuming $\beta_i \sim \beta_e$. In the electrostatic limit, the Alfv\'en transit frequency is much higher than the ITG growth rate. As either $\beta_e$ or $R$ increases, however, the two timescales become more comparable, signalling the onset of stronger electromagnetic effects. Consequently, the linear scaling $r_{\mathrm{MR}}\propto\beta_e$ breaks down at smaller values of $\beta_e$ for larger aspect ratios, consistent with the observed dependence of $\beta_{e,\mathrm{sb}}$ on $R$.

\subsubsection{Predicting the non-zonal-transition boundary using low-$\beta_e$ simulations}
\label{sec:predict}
In this section, we develop a simple practical prescription for predicting the
non-zonal-transition boundary from simulations performed well below the
transition. As demonstrated in
Fig.~\ref{fig:q_beta_plots}, the onset of the non-zonal transition
coincides with the Maxwell stress becoming comparable to the Reynolds
stress, corresponding to a transfer ratio of
$r_{\mathrm{MR}}\sim1$. Parameterising the linear scaling law (\ref{eq:transfer_ratio_scaling_estimate}) as
\begin{equation}
\label{eq:linear_scaling_parameterised}
r_{\mathrm{MR}}=\alpha_1 q^2\beta_e\frac{R}{L_{T_i}},
\end{equation}
where $\alpha_1$ is a proportionality constant independent of $q$, $\beta_e$, or $R/L_{T_i}$, one could perform a single simulation at a conveniently (from the numerical point of view) low $\beta_e$,
measure $r_{\mathrm{MR}}$, determine $\alpha_1$, and predict the transition boundary by setting $r_{\mathrm{MR}} = 1$, so
\begin{equation}
\label{eq:wrong_estimate}
(q^2\beta_e)_{\mathrm{critical}}=C_{\mathrm{nl}},\quad\mathrm{where}
\quad
C_{\mathrm{nl}}=\left(\alpha_1\frac{R}{L_{T_i}}\right)^{-1}.
\end{equation}
Such a procedure would eliminate the need for a parameter scan and expensive simulations in
the vicinity of the transition.

In practice, however, as we have seen in Section~\ref{sec:numerical_validation}, $r_{\mathrm{MR}}$ begins to deviate from the scaling (\ref{eq:linear_scaling_parameterised}) noticeably before reaching $r_{\mathrm{MR}} = 1$, increasing much
more rapidly as the transition is approached. Consequently, (\ref{eq:wrong_estimate}) would significantly
overestimate the critical value of $q^2\beta_e$. Therefore, we introduce a modelling threshold
\begin{equation}
\label{eq:correct_estimate}
r_{\mathrm{MR}}=\alpha_2 < 1,\quad \mathrm{whence}\quad C_{\mathrm{nl}} = \alpha_2\left(\alpha_1\frac{R}{L_{T_i}}\right)^{-1}.
\end{equation}
In principle, $\alpha_2$ could depend on the equilibrium parameters such as $R/L_{T_i}$ or geometry. As shown in Figs.~\ref{fig:transfer_ratio_beta} and~\ref{fig:transfer_ratio_beta_st40}, the scaling law departs from its linear behaviour at lower values of $\beta_e$ (and lower values of $r_{\mathrm{MR}}$) for larger major radius $R$, after which $r_{\mathrm{MR}}$ increases more rapidly than linearly. Since the purpose of introducing the empirical parameter $\alpha_2$ is to avoid overestimating the non-zonal-transition threshold when using the linear scaling, the results of Figs.~\ref{fig:transfer_ratio_beta} and~\ref{fig:transfer_ratio_beta_st40} suggest that smaller values of $\alpha_2$ should be adopted for larger values of $R/L_{T_i}$. We now seek a plausible dependence of $\alpha_2$ on $R/L_{T_i}$ based on the available CBC and ST40 simulation data.

\begin{table}
\centering
\small
\renewcommand{\arraystretch}{1.4}
\begin{tabular}{|c|c|c|c|c|c|}
\hhline{|------|}
Case & $C_{\mathrm{nl}}$ & $R/L_{T_i}$ & $\alpha_1$ & $\alpha_2$ & $\alpha_2 R/L_{T_i}$  \\
\hhline{|------|}
CBC  & $0.014$ & $6.90$ & $0.646$ & $0.0624$ & $0.430$\\
ST40 & $0.049$ & $3.80$ & $0.744$ & $0.139$ & $0.527$\\
\hhline{|------|}
\end{tabular}
\caption{Summary of the best-fit $C_{\mathrm{nl}}$ (taken from Fig.~\ref{fig:q_beta_plots}), $R/L_{T_i}$, $\alpha_1$, and $\alpha_2$ for the CBC and ST40 cases. The values of $\alpha_1$ are evaluated for the representative low-$\beta_e$ simulations with $(q,\beta_e,\tilde{R})=(1.4,\,10^{-5},\,3.0)$ for CBC and $(q,\beta_e,\tilde{R})=(1.7,\,7.5\times10^{-6},\,1.82)$ for ST40, with the rest of the equilibrium parameters the same as those in Fig~\ref{fig:q_beta_plots}.}
\label{tab:Cnl_summary}
\end{table}

The fitted parameters are summarised in Table~\ref{tab:Cnl_summary}. The best-fit values of $C_{\mathrm{nl}}$ are obtained from the empirical non-zonal-transition boundaries (dashed black lines) shown in Fig.~\ref{fig:q_beta_plots}, while $\alpha_1$ is determined from the representative low-$\beta_e$
simulations listed in the table using the linear scaling law (\ref{eq:linear_scaling_parameterised}). The corresponding values of $\alpha_2$ are then obtained from~(\ref{eq:correct_estimate}). Although the inferred values of $\alpha_2$ differ by more than a factor of two between the CBC and ST40 cases, the product $\alpha_2R/L_{T_i}$ is remarkably similar. This suggests that the principal dependence of $\alpha_2$ may be captured through $R/L_{T_i}$ via an inverse relationship, motivating the conjecture
\begin{equation}
\label{eq:alpha2_conjecture}
\alpha_2\frac{R}{L_{T_i}}\approx 0.5,
\end{equation}
which, when substituted into (\ref{eq:correct_estimate}), gives
\begin{equation}
\label{eq:final_Cnl_expression}
C_{\mathrm{nl}}\approx\frac{0.5}{\alpha_1(R/L_{T_i})^2}.
\end{equation}
The resulting expression can be regarded as a practical empirical prescription for estimating the non-zonal-transition boundary from a single low-$\beta_e$ nonlinear simulation required to determine $\alpha_1$. It should be emphasised, however, that a systematic scan in $R/L_{T_i}$ is required to validate the conjecture (\ref{eq:alpha2_conjecture}) and thereby place (\ref{eq:final_Cnl_expression}) on a firmer footing.

\section{Conclusion}
\label{sec:conclusion}
We have studied the role of the Maxwell and Reynolds stresses in determining the transition to a high-transport turbulent state (the non-zonal transition), and introduced a method that enables us, with minimal computational effort, to predict the parameters at which the transition occurs.

Gyrokinetic simulations of the CBC and ST40 equilibria show that, when the effect of the kinetic Maxwell stress on the zonal-flow dynamics becomes comparable to that of the kinetic Reynolds stress, this results in an electromagnetically turbulent state with weak zonal flows and extreme heat fluxes. The kinetic Maxwell stress extracts energy from zonal flows and feeds it into turbulence, negating the energy injection by the kinetic Reynolds stress into the zonal flows. This physical picture was also present in the fluid limit of gyrokinetics discussed in Appendix~\ref{sec:fluid-limit} and first explored in \cite{zhang_2026}, whose conclusions thus qualitatively carry over to the more realistic tokamak models. We have quantified the influences of the kinetic Maxwell and Reynolds stresses on the zonal flow and turbulence saturation using the time-averaged, large-scale energy-transfer functions (Section~\ref{sec:turbulent energy transfers}). The Maxwell and Reynolds stresses become comparable near the non-zonal transition, corresponding to a Maxwell-to-Reynolds energy-transfer ratio $r_{\mathrm{MR}}\sim1$, and the transition boundary is well described by
$q^2\beta_e=C_{\mathrm{nl}}$, where $C_{\mathrm{nl}}$ depends on the geometry of the tokamak configuration and the ion temperature gradient.

We have shown analytically that, in the low-$\beta_e$ limit, $r_{\mathrm{MR}}$ satisfies a simple scaling~(\ref{eq:transfer_ratio_scaling_estimate}), and verified it using gyrokinetic simulations for both CBC and ST40 representative equilibria. As the non-zonal transition is approached, however, $r_{\mathrm{MR}}$ departs systematically from the linear scaling and increases more rapidly with $\beta_e$. This effect is currently not understood theoretically and appears stronger in configurations with larger aspect ratios. In Section~\ref{sec:predict}, we proposed a simple extrapolation (\ref{eq:final_Cnl_expression}) requiring a single parameter that can be measured from a single low-$\beta_e$ simulation, avoiding expensive runs close to the transition, where electromagnetic effects impose a much more restrictive time-step constraint.

In the cases that we investigated, even though the KBM and the lower IBM thresholds approximately follow the same scaling trend, viz., $q^2\beta_e = C_{\mathrm{KBM}}$ and $q^2\beta_e = C_{\mathrm{IBM}}$, the corresponding constants are found to satisfy $C_{\mathrm{nl}} < C_{\mathrm{KBM}} < C_{\mathrm{IBM}}$. This suggests that the linear thresholds of electromagnetic instabilities such as KBMs do not coincide with the non-zonal-transition boundary. Therefore, from a practical standpoint, when determining the operational beta limits for a given device, it is important to account for the constraints imposed by micro-scale turbulence, rather than relying solely on linear-stability thresholds, i.e., because of the nonlinear energy-transfer processes inherent to micro-scale turbulence, the operational beta limit may be more restrictive than suggested by linear-stability analyses. 

The additional studies presented in Appendices~\ref{sec:large_scale} and~\ref{sec:long_evolution_and_transport_hysteresis} further support the robustness of the stress-based picture. In Appendix~\ref{sec:large_scale}, we show that, although the detailed high-$k_x$ transfer spectra are sensitive to numerical resolution, the large-scale energy transfers $T_{\phi}$ and $T_{\apar}$ that determine $r_{\mathrm{MR}}$ remain essentially unchanged, thereby justifying our focus on the large-scale stress competition. We also demonstrate that toroidal secondary modes (TSMs), whose dynamics are likewise affected by the competition between $T_{\phi}$ and $T_{\apar}$, are present in the electromagnetic simulations. Unlike in the electrostatic cases reported by \cite{nies_2024}, however, the TSMs in our simulations remain subdominant to the stationary zonal flows, which is likely a consequence of the inclusion of kinetic electrons. In Appendix~\ref{sec:long_evolution_and_transport_hysteresis}, we show that simulations in larger domains may exhibit slow growth of the box-size zonal mode, which, after a long initial transient, can suppress Maxwell stress and enable the system to saturate with a modest heat flux even at some values of $q^2\beta_e$ at which it would blow up if initialised with small noise, i.e., a transport hysteresis is possible~\citep{Rath_2022}. Despite these long-time effects, the large-scale transfer ratio $r_{\mathrm{MR}}$ remains a useful indicator of proximity to the non-zonal transition, provided that it is evaluated during the transient period before the onset of the slow evolution of the box-scale zonal mode. During this transient period, the competition between the Reynolds and Maxwell stresses is strongest and determines whether there is a danger of the turbulence accessing an EM high-transport state.
 
A key direction for future work is to understand the physical origin of the nonlinear departure from the scaling (\ref{eq:transfer_ratio_scaling_estimate}) and the form and origin of the $R/L_{T_i}$ dependence of the transition threshold (\ref{eq:final_Cnl_expression}). Another important direction is to test the proposed prediction method in equilibria derived from other devices in order to assess its generality across a broader range of plasma configurations.

\section*{Acknowledgements}
The work of Y.~Z. was supported in part by Tokamak Energy Ltd. His work, and the work of A.~A.~S., was also supported in part by the Simons Foundation via a Simons Investigator Award to A.~A.~S. The work of M.~B.,  A.~A.~S. and P.~G.~I. was supported in part by EPSRC (grant EPR034737/1). This work was also part-funded by the EPSRC Fusion Grant 2022/27 (grant number EP/W006839/1). The work of T.A. was supported in part by the Laboratory Directed Research and Development (LDRD) Program at the Princeton Plasma Physics Laboratory for the U.S. Department of Energy under Contract No. DE-AC02-09CH11466. The United States Government retains a non-exclusive, paid-up, irrevocable, world-wide license to publish or reproduce the published form of this manuscript, or allow others to do so, for United States Government purposes.

The authors would like to thank Felix Parra, Richard Nies, Michele Romanelli, and the rest of the Oxford plasma theory group for fruitful discussions and useful comments.

For the purpose of open access, the author(s) has applied a Creative Commons Attribution (CC BY) licence (where permitted by UKRI, ‘Open Government Licence’ or ‘Creative Commons Attribution No-derivatives (CC BY-ND) licence’ may be stated instead) to any Author Accepted Manuscript version arising.

\bibliographystyle{jpp}

\bibliography{bib}

\appendix 
\section{Derivation of evolution equation for zonal flows}
\label{app:derivation_details}
In this appendix, we derive the evolution equation (\ref{eq:phi_zonal_evo2}) for zonal flows. Our starting point is (\ref{eq:phi_zonal_evo1}), which we reproduce here for convenience:
\begin{equation}
\label{eq:phi_zonal_evo1_appendix}
 \frac{\partial }{\partial t}\sum_s  n_s \frac{Z^2_s e}{T_s} \avg{\phi }{yz} =  \sum_s Z_s \avg{\int \sd^3 \bm{v}\,\frac{\partial \gyror{h_s}}{\partial t}}{yz}.
\end{equation}
We proceed by replacing the right-hand side of (\ref{eq:phi_zonal_evo1_appendix}) using the gyroaveraged version of the gyrokinetic equation (\ref{eq:gk_eq}), integrating over the velocity space and averaging over a flux surface:
\begin{align}
\label{eq:phi_zonal_evo1_appendix_substituted}
& \frac{\partial }{\partial t}\sum_s n_s\frac{Z^2_s e}{T_s}  \avg{\phi }{yz} =  - \underbrace{\sum_s Z_s\avg{ \int \sd^3 \bm{v}\,\vpar (\unit{b}\cdot\grad z)\frac{\partial \gyror{h_s}}{\partial z}}{yz}}_{\circled{1}} \nonumber \\ &-\underbrace{\sum_s Z_s \avg{\int \sd^3 \bm{v}\,\bm{v}_{Ms} \cdot \gradperp \gyror{h_s}}{yz}}_{\circled{2}} + \underbrace{\sum_s Z_s \avg{\int \sd^3 \bm{v}\,\frac{\mu}{m_s} (\unit{b}\cdot\grad B)\frac{\partial \gyror{h_s}}{\partial \vpar}}{yz}}_{\circled{3}} \nonumber \\ &- \underbrace{\sum_s Z_s \avg{\int \sd^3 \bm{v}\,\gyror{\gyroRs{\vchi}\cdot \gradperp h_s}}{yz}}_{\circled{4}} - \underbrace{\sum_s Z_s \avg{\int \sd^3 \bm{v}\,\gyror{\gyroRs{\vchi}}\cdot \grad F_s}{yz}}_{\circled{5}}\nonumber \\ & = \underbrace{\sum_s \frac{Z^2_s e}{T_s} \avg{\int \sd^3\bm{v}\, F_s\frac{\partial \gyror{\gyroRs{\chi}}}{\partial t}}{yz}}_{\circled{6}}   + \sum_s Z_s \avg{\int \sd^3\bm{v}\, \gyror{\gyroRs{C\left[h_s\right]}} }{yz}.
\end{align}
Let us now calculate the right-hand side of this equation term by term. 

\noindent \textbf{Term} \circled{1} : This term, which contains the parallel-streaming contribution, is zero because
\begin{align}
\label{eq:parallel_stream_flux_ave_v_int}
\avg{\int \sd^3 \bm{v} \,\vpar (\unit{b}\cdot\grad z) \frac{\partial \gyror{h_s}}{\partial z}}{yz} & \propto  \int \sd y\,\sd z\,\frac{\unit{b}\cdot\grad z}{|\bm{B}\cdot\grad z|}\int \sd \vpar\,\sd \mu \frac{2\pi B}{m_s}  \,\vpar \frac{\partial \gyror{h_s}}{\partial z} \nonumber \\ & \propto \int \sd y\,\sd z \int \sd \vpar\,\sd \mu  \,\vpar \frac{\partial \gyror{h_s}}{\partial z} \nonumber \\ & \propto \int  \sd z\,\frac{\partial}{\partial z} \left( \int \sd y \int \sd \vpar\,\sd\mu \, \vpar  \gyror{h_s}\right),
\end{align}
where we have used the expression for the velocity integral
\begin{equation}
\label{eq:vpamu_int}
\int \sd^3 \bm{v} = \int^{2\pi}_0 \sd \vartheta \int^{\infty}_{-\infty} \sd \vpar \int^{\infty}_0 \frac{B}{m_s} \sd\mu,
\end{equation} 
with $\vartheta$ the gyro-angle, as well as (\ref{eq:flux_ave_yz}) and the fact that $ (\sd y/\sd \alpha) (\sd x/\sd \psi)$ is constant along the field line. In general, fluctuations need not be periodic in $z$ \citep{Beer_1995}: magnetic shear leads to shearing of the flux tube when moving along the magnetic field. However, this shearing is smeared out upon averaging over $y$. Consequently, $y$-averaged terms like (\ref{eq:parallel_stream_flux_ave_v_int}) are periodic in $z$. Hence, the $z$-integral evaluates to zero.

\noindent \textbf{Term} \circled{2} : The magnetic-drift term can be decomposed into its radial and binormal components, viz.,
\begin{equation}
\bm{v}_{Ms}\cdot \gradperp \gyror{h_s} = (\bm{v}_{Ms}\cdot\grad x) \frac{\partial \gyror{h_s}}{\partial x} + (\bm{v}_{Ms}\cdot\grad y) \frac{\partial \gyror{h_s}}{\partial y}.
\end{equation}
It is easy to see that the binormal drift vanishes after flux-surface averaging by using (\ref{eq:flux_ave_yz}) and noting that $\gyror{h_s}$ is periodic in $y$ and $\bm{v}_{Ms}\cdot\grad y$ is itself independent of $y$. Consequently, the magnetic-drift contribution becomes
\begin{equation}
\label{eq:term2}
\sum_s Z_s \avg{\int \sd^3 \bm{v}\,\bm{v}_{Ms} \cdot \gradperp \gyror{h_s}}{yz} = \sum_s Z_s \avg{\int \sd^3 \bm{v}\, (\bm{v}_{Ms}\cdot\grad x) \frac{\partial \gyror{h_s}}{\partial x}}{yz}.
\end{equation}

\noindent \textbf{Term} \circled{3} : The mirror contribution contains a total derivative of $\vpar$. It thus vanishes under integration over $\vpar$, assuming that $h_s$ decays to zero as $\vpar \to \infty$.

\noindent \textbf{Term} \circled{4} : We rewrite the nonlinearity in terms of Poisson brackets:
\begin{equation}
\gyroRs{\vchi}\cdot \gradperp h_s = \frac{c}{B}\left[\unit{b}\times \left(\grad x \frac{\partial \gyroRs{\chi}}{\partial x} + \grad y \frac{\partial \gyroRs{\chi}}{\partial y}\right)\right]\cdot \left[(\grad x) \frac{\partial h_s}{\partial x} + (\grad y) \frac{\partial h_s}{\partial y}\right],
\end{equation}
which leads to
\begin{align}
\label{eq:term4}
& \sum_s Z_s \avg{\int \sd^3 \bm{v}\,\gyror{\gyroRs{\vchi}\cdot \gradperp h_s}}{yz} \nonumber \\ & = \sum_s Z_s \avg{\frac{c}{B}\unit{b}\cdot (\grad x \times \grad y) \int \sd^3 \bm{v}\,\gyror{\left \lbrace \gyroRs{\phi}, h_s \right \rbrace}}{yz}\nonumber \\ & - \sum_s Z_s \avg{\frac{1}{B}\unit{b}\cdot (\grad x \times \grad y)\int \sd^3 \bm{v}\,\vpar \gyror{\left \lbrace \gyroRs{ \apar}, h_s \right \rbrace}}{yz} \nonumber\\ & -  \sum_s Z_s \avg{\frac{1}{B}\unit{b}\cdot (\grad x \times \grad y)\int \sd^3 \bm{v}\, \gyror{\left \lbrace \gyroRs{\bm{v}_{\perp}\cdot\aperp}, h_s \right \rbrace}}{yz},
\end{align}
where we have used (\ref{eq:vchi_definition}), (\ref{eq:chi_definition}), and (\ref{eq:gradperp_definition}).

\noindent \textbf{Term} \circled{5} : We rewrite the free-energy-injection term using flux-tube coordinates:
\begin{align}
\label{eq:vchi_dot_Fs}
\gyror{\gyroRs{\vchi}} \cdot \grad F_s \nonumber & = \frac{c}{B}\left[\frac{(\grad y \times \grad x)}{|\grad x||\grad y|}\times(\grad x) \frac{\partial \gyror{\gyroRs{\chi}}}{\partial x}\right]\cdot (\grad x)  \frac{\partial F_s}{\partial x} \nonumber\\& +  \frac{c}{B}\left[\frac{(\grad y \times \grad x)}{|\grad x||\grad y|}\times(\grad y) \frac{\partial \gyror{\gyroRs{\chi}}}{\partial y}\right] \cdot(\grad x)\frac{\partial F_s}{\partial x},
\end{align}
where we have used $\unit{b} = \grad y \times \grad x/(|\grad x||\grad y|)$ and $\grad F_s = (\grad x) \partial F_s/\partial x$.
The first term on the right-hand side of (\ref{eq:vchi_dot_Fs}) is clearly zero. The second term can be written as a total derivative of $y$ since only $\gyror{\gyroRs{\chi}}$ varies in $y$. Hence, the free-energy-injection vanishes under flux-surface averaging. 

\noindent \textbf{Term} \circled{6} : 
\begin{align}
\label{eq:term6}
&\sum_s \frac{Z^2_s e}{T_s} \avg{\int \sd^3\bm{v}\, F_s\frac{\partial \gyror{\gyroRs{\chi}}}{\partial t}}{yz} = \frac{\partial}{\partial t}\sum_s \frac{Z^2_s e}{T_s} \avg{\int \sd^3\bm{v}\, F_s \gyror{\gyroRs{\phi}}}{yz}\nonumber \\ & - \frac{\partial}{\partial t}\sum_s \frac{Z^2_s e}{c T_s} \avg{\int \sd^3\bm{v}\,\vpar F_s\gyror{\gyroRs{\apar}}}{yz} - \frac{\partial}{\partial t}\sum_s \frac{Z^2_s e}{c T_s} \avg{\int \sd^3\bm{v}\, F_s \gyror{\gyroRs{\bm{v}_{\perp}\cdot\aperp}}}{yz}.
\end{align}
The second term on the right-hand side of this equation, containing the inductive electric field, is zero because the integrand is odd in $\vpar$.

Collecting all the non-vanishing terms from (\ref{eq:term2}), (\ref{eq:term4}), (\ref{eq:term6}), the collisional contribution, and writing $n_s$ on the left-hand side of (\ref{eq:phi_zonal_evo1_appendix_substituted}) as $\int \sd^3\bm{v}\,F_s$, we obtain the final form of the zonal-flow-evolution equation (\ref{eq:phi_zonal_evo2}). 

\section{Reynolds-Maxwell stress competition in the fluid limit}
\label{sec:fluid-limit}
In this appendix, we show analytically that the Reynolds and Maxwell stresses have opposite effects on the zonal flow in the fluid limit. The starting point is \eqref{eq:dzonal_dt}, from which we will simplify the expressions for the stresses on the right-hand side. 

\subsection{A fluid equation for zonal flows}
We consider long-wavelength fluctuations and conduct a local analysis at $z=0$. This is an ad-hoc approximation motivated primarily by a desire to simplify analysis but also by the fact that turbulence is driven most strongly in the outboard midplane. In this approximation, the flux-surface average becomes an average over a portion of the flux tube in the vicinity of $z = 0$, which is effectively an average over $y$:
\begin{equation}
\label{eq:flux_ave_yz_z=0}
\avg{g (\bm{r})}{yz} \rightarrow \avg{ g(z=0,x,y)}{y}.
\end{equation}
At $z=0$, we also assume that the vectors $\unit{b}$, $\grad x$ and $\grad y$ form an orthonormal basis that satisfies $|\grad x| = |\grad y| = 1$ and $\unit{b}\cdot (\grad x \times\grad y) = -1$, which again can be realised for large-aspect-ratio, circular flux surfaces. Note that in this case $\kperp^2 = k^2_x + k^2_y$. 

First, consider the expression (\ref{eq:phi_stress}) for $\Pi_{\phi}$. To the lowest order in $\kperp\rho_i$, $\Pi_{\phi}$ vanishes. This is because $\sum_s Z_s \int \sd^3 \bm{v}\,h_s(\bm{r}) \propto \phi$ due to quasineutrality \eqref{eq:QN} and $\left\lbrace \phi, \phi \right\rbrace = 0$. Note that $\kperp\rho_e \ll \kperp\rho_i \ll 1$, which means we shall retain next-order finite-Larmor-radius (FLR) corrections for the ions, while keeping only the lowest-order terms for the electrons. The ion contribution to the Poisson bracket appearing in $\Pi_{\phi}$ is then approximately
\begin{align}
\label{eq:phi_poisson}
& \avg{\left\lbrace \avg{\phi}{\bm{R}_i}, h_i\right\rbrace}{\bm{r}}  \approx \sum_{\bm{k}_{\perp},\bm{k}_{\perp}^{\prime}} e^{i\bm{k}_{\perp}\cdot\bm{r} + i\bm{k}_{\perp}^{\prime}\cdot\bm{r}} \left(-k_xk'_y + k_yk'_x\right) \Bigg[1 - \bm{k}_{\perp}\bm{k}_{\perp}^{\prime}:\left(\frac{1}{2}\frac{v^2_{\perp}\rho^2_i}{v^2_{\mathrm{th}i}}\textbf{I}\right) \nonumber\\ & - \frac{1}{4}\frac{v^2_{\perp}\rho^2_i}{v^2_{\mathrm{th}i}}k_{\perp}^{\prime 2} - \frac{1}{2}\frac{v^2_{\perp}\rho^2_i}{v^2_{\mathrm{th}i}}k^2_{\perp}\Bigg]\phi_{\bm{k}_{\perp}}h_{i,\bm{k^{\prime}}_{\perp}}\nonumber \\& =  \left\lbrace \phi, \left(1 + \frac{1}{4}\frac{\vperp^2\rho^2_i}{\vthi^2}\gradperp^2\right) h_i\left(\bm{r}\right) \right\rbrace + \frac{1}{2} \frac{v^2_{\perp}\rho^2_i}{v^2_{\mathrm{th}i}} \textbf{I}:\left\lbrace \gradperp \phi, \gradperp h_i\left(\mathbf{r}\right) \right\rbrace + \frac{1}{2}\frac{v^2_{\perp}\rho^2_i}{v^2_{\mathrm{th}i}}\left\lbrace \nabla^2_{\perp}\phi, h_i\left(\mathbf{r}\right)\right\rbrace,
\end{align}
where we have expanded $e^{i\bm{k}_{\perp}\cdot \boldsymbol{\rho}}$, with $\boldsymbol{\rho} = \bm{r} - \bm{R}_i$, and the Bessel function $J_0(\kperp \rho_i \vperp/\vthi)$ contained in $\avg{\phi}{\bm{R}_i}$ in $\kperp \rho_i \ll 1$. The identity matrix $\textbf{I}$ is defined as $\hat{\bm{x}}\hat{\bm{x}} + \hat{\bm{y}}\hat{\bm{y}}$. We also used the following properties of the gyroaverage:
\begin{equation}
\label{eq:gyro_property}
\avg{\boldsymbol{\rho}}{\mathbf{r}} = 0,\,\,\,  \avg{\boldsymbol{\rho}\boldsymbol{\rho}}{\bm{r}} = \frac{1}{2}\frac{\vperp^2\rho^2_i}{\vthi^2}\textbf{I}.
\end{equation}
The Poisson bracket from the electron contribution to $\Pi_{\phi}$, to lowest order in $\kperp \rho_e$, is
\begin{equation}
\label{eq:phi_poisson_electrons}
\avg{\left\lbrace \avg{\phi}{\bm{R}_e}, h_e\right\rbrace}{\bm{r}} \approx \left\lbrace \phi, h_e(\bm{r})\right\rbrace.
\end{equation}
Using (\ref{eq:phi_poisson}) and (\ref{eq:phi_poisson_electrons}), we rewrite $\Pi_{\phi}$ (\ref{eq:phi_stress}) as
\begin{align}
\label{eq:Pi_phi_fluid_pre}
\Pi_{\phi} = \frac{c}{B}\avg{ \left\lbrace \phi, \left[Z_i\int \sd^3 \bm{v}\,\left(1 + \frac{1}{4}\frac{\vperp^2\rho^2_i}{\vthi^2}\gradperp^2\right)h_i(\bm{r}) - \int \sd^3 \bm{v} \,h_e(\bm{r})\right]\right\rbrace}{y} \nonumber \\ + \frac{Z_i c\rho^2_i}{2 B}\avg{\textbf{I}:\left\lbrace\gradperp\phi, \gradperp \int \sd^3 \bm{v}\,\frac{\vperp^2}{\vthi^2} h_i(\bm{r}) \right\rbrace}{y} + \frac{Z_i c\rho^2_i}{2B}\avg{\left\lbrace\gradperp^2\phi, \int \sd^3 \bm{v}\,\frac{\vperp^2}{\vthi^2}h_i(\bm{r}) \right\rbrace}{y}.
\end{align}
The first term on the right-hand side is zero due to the expanded quasineutrality constraint
\begin{equation}
\label{eq:expanded_QN}
Z_i\int \sd^3 \bm{v}\,\left(1 + \frac{1}{4}\frac{\vperp^2\rho^2_i}{\vthi^2}\gradperp^2\right)h_i(\bm{r}) - \int \sd^3 \bm{v} \,h_e(\bm{r}) = \left(\frac{Z^2_in_i}{T_i} + \frac{n_e}{T_e}\right)e\phi.
\end{equation}
The last two terms of (\ref{eq:Pi_phi_fluid_pre}) can be evaluated using the fact that
\begin{equation}
h_i(\bm{r}) = \delta f_i(\bm{r}) + \frac{Z_ie\phi}{T_i}F_i(\bm{r})\quad \mathrm{and}\quad \frac{1}{n_i}\int \sd^3 \bm{v}\,  \frac{\vperp^2}{\vthi^2}\delta f_i(\bm{r}) = \frac{\delta p_{\perp,i}}{p_i}.
\end{equation}
The resulting expression for $\Pi_{\phi}$ is
\begin{align}
\label{eq:Pi_phi_fluid}
\Pi_{\phi} & = -\frac{Z_in_i\rho_ic^2}{\vthi B^2}\avg{\left\lbrace \phi, \gradperp^2\phi \right\rbrace}{y} + \frac{Z_i n_ic\rho^2_i}{2B}\avg{\gradperp\cdot\llbrace \gradperp \phi, \frac{\delta p_{\perp,i}}{p_i} \rrbrace}{y} \nonumber \\ &  + O\left( n_i\rho^4_i\gradperp^4\frac{c\gradperp^2\phi }{B}\right).
\end{align}
After evaluating the reduced flux-surface average \eqref{eq:flux_ave_yz_z=0} and integrating by parts, \eqref{eq:Pi_phi_fluid} becomes:
\begin{equation}
\label{eq:pi_phi_Z}
\Pi_{\phi} =  \frac{Z_in_i\rho_ic^2}{\vthi B^2}\frac{\partial^2}{\partial x^2}\avg{\frac{\partial \phi^{\prime}}{\partial x} \frac{\partial \phi^{\prime}}{\partial y}}{y} + \frac{Z_in_ic\rho^2_i}{2B}\frac{\partial^2}{\partial x^2}\avg{\frac{\partial \phi^{\prime}}{\partial x} \frac{\partial}{\partial y}\left(\frac{\delta p'_{\perp, i}}{p_i}\right)}{y},
\end{equation}
where the prime denotes the non-zonal part of the field, viz., $\phi^{\prime} = \phi - \avg{\phi}{y}$. 

The kinetic Maxwell stress $\Pi_{\apar}$ (\ref{eq:apar_stress}) can be simplified in an analogous way. Consider the lowest-order contribution to $\Pi_{\apar}$ in $\kperp \rho_i \ll 1$:
\begin{align}
\label{eq:pi_apar_Z}
\Pi_{\apar} & = - \frac{1}{B}\avg{\left\lbrace \apar, \sum_s Z_s\int \sd^3 \bm{v}\,\vpar h_s(\bm{r})\right\rbrace}{y} + O\left[\rho^2_i \gradperp^2\frac{c(\gradperp^2\apar)^2}{4\pi e B}\right] \nonumber \\ & = - \frac{c}{4\pi e B}\frac{\partial^2}{\partial x^2}\avg{\frac{\partial \apar^{\prime}}{\partial x}\frac{\partial \apar^{\prime}}{\partial y}}{y} + O\left[\rho^2_i \gradperp^2\frac{c(\gradperp^2\apar)^2}{4\pi e B}\right],
\end{align}
where we have used the parallel component of Amp\`ere’s law \eqref{eq:parallel_ampere} and the approximation~\eqref{eq:flux_ave_yz_z=0}. We will primarily focus on the competition between the Reynolds, diamagnetic, and Maxwell stresses. Hence, we will ignore all other contributions to the right-hand side of the zonal-flow-evolution equation \eqref{eq:dzonal_dt}. Using \eqref{eq:pi_phi_Z} and \eqref{eq:pi_apar_Z}, we express \eqref{eq:dzonal_dt} as
\begin{equation}
\label{eq:zonal_flow_fluid}
\frac{\partial v^{\rm{Z}}}{\partial t} + \frac{\partial}{\partial x}\avg{v^{\prime}_x v^{\prime}_y}{y} + \frac{\rho_i \vthi}{2}\frac{\partial}{\partial x}\avg{v^{\prime}_y \frac{\partial}{\partial y}\frac{\delta p^{\prime}_{\perp,i}}{p_i}}{y} - \frac{v^2_A}{B^2}\frac{\partial }{\partial x}\avg{\delta B^{\prime}_x \delta B^{\prime}_y}{y} = 0,
\end{equation}
where $v_A = B/\sqrt{4\pi m_i n_i}$ is the Alfv\'{e}n speed, $v_x = (c/B)\partial \phi/\partial y$ and \mbox{$v_y = -(c/B)\partial \phi/\partial x$} are the $x$ and $y$ components of the $\bm{E}\times\bm{B}$ flow $\bm{v}_{E} = v_x \unit{x} + v_y \unit{y}$ and $\delta B_x = - \partial \apar/\partial y$ and $\delta B_y = \partial \apar/\partial x$ are the $x$ and $y$ components of the perturbed magnetic field $\delta\bm{B}_{\perp} = \delta B_x \unit{x} + \delta B_y \unit{y}$. Note that, to obtain (\ref{eq:zonal_flow_fluid}) from  (\ref{eq:dzonal_dt}), one needs to integrate both sides of (\ref{eq:zonal_flow_fluid}) over $x$. The integration constant is zero as $\phi$ is a smooth and periodic function of $x$.

The second and third terms in \eqref{eq:zonal_flow_fluid} are derivatives of the fluid Reynolds and diamagnetic stresses, which were found previously in fluid models of ITG turbulence \citep{ivanov_2020, ivanov_2022} to drive and damp zonal flows, respectively. The fourth term is the derivative of the fluid Maxwell stress \citep{zhang_2026}. Inspecting \eqref{eq:zonal_flow_fluid}, one finds that, even in the absence of any possible damping caused by the diamagnetic stress, the zonal-flow drive still vanishes for Alfvénic fluctuations, because for Alfvén waves we have $\delta \bm{B}_{\perp}/B = \bm{v}_{E}/v_A$, leading to a complete cancellation between the fluid Reynolds and Maxwell stresses. However, as we will show in Appendix~\ref{sec:stress_compet_zonal_dominate}, the partial or complete cancellation between the two stresses does not only occur for Alfvénic turbulence, but also occurs whenever the turbulence is dominated by zonal flows.

\subsection{Reynolds-Maxwell stress competition in zonal-flow-dominated turbulence}
\label{sec:stress_compet_zonal_dominate}
Here we consider a regime where the turbulence is strongly sheared by zonal flows that evolve on a much longer time scale than the non-zonal fluctuations. We assume that the zonal flows produce shearing zones of characteristic width $l_x$ along $x$, within which the zonal shear $\gamma_E = \partial v^{\rm{Z}}/\partial x$ is approximately constant. We further assume that $l_x$ is large compared with the perpendicular spatial scale of the non-zonal fluctuations. As a result, the non-zonal fluctuations are treated as statistically periodic inside a shearing zone.

Multiplying (\ref{eq:zonal_flow_fluid}) by $v^{\rm{Z}}$, integrating over $x$ within the shearing zone and dropping the diamagnetic-stress term for simplicity yields an evolution equation for zonal-flow energy:
\begin{equation}
\label{eq:dvz2_dt_shearing_zone}
\frac{\partial }{\partial t}\overline{\frac{(v^{\rm{Z}})^2}{2}} = \gamma_E\overline{\avg{v^{\prime}_x v^{\prime}_y}{y}} - \frac{v^2_A}{B^2} \gamma_E\overline{\avg{\delta B^{\prime}_x \delta B^{\prime}_y}{y}},
\end{equation}
where the overline denotes averaging across the shearing zone; we have also used integration by parts and the fact that the shear $\gamma_E$ is constant inside the zone. The Reynolds contribution $\gamma_E\overline{\avg{v^{\prime}_x v^{\prime}_y}{y}}$ and the Maxwell contribution $(v^2_A/B^2) \gamma_E \overline{\avg{\delta B^{\prime}_x \delta B^{\prime}_y}{y}}$ can be shown to be sign-definite. To see that, we first express the right-hand side of~(\ref{eq:dvz2_dt_shearing_zone}) in Fourier space:
\begin{equation}
\label{eq:dvz2_dt_shearing_zone_kspace}
\frac{\partial }{\partial t}\overline{\frac{(v^{\rm{Z}})^2}{2}}  = - \frac{c^2}{B^2} \gamma_E \sum_{\bm{k}_{\perp}} k_xk_y|\phi'_{\bm{k}_{\perp}}|^2 + \frac{v^2_A}{B^2}\gamma_E\sum_{\bm{k}_{\perp}} k_xk_y |A'_{\parallel,\bm{k}_{\perp}}|^2.
\end{equation}
In the shearing zone, the $k_x$ and $k_y$ of the non-zonal fluctuation $\phi'$ are not independent. For fluctuations that are stationary in the shearing frame, one finds $k_x \approx - \gamma_E t k_y$ in the laboratory frame, valid when $\gamma_E t \gg 1$. Substituting this expression for $k_x$ into (\ref{eq:dvz2_dt_shearing_zone_kspace}) yields
\begin{equation}
\frac{\partial }{\partial t}\overline{\frac{(v^{\rm{Z}})^2}{2}} \approx \frac{c^2}{B^2}\gamma_E^2 t \left(\sum_{\bm{k}_{\perp}} k^2_y|\phi'_{\bm{k}_{\perp}}|^2 - \frac{v^2_A}{c^2}\sum_{\bm{k}_{\perp}} k^2_y |A'_{\parallel,\bm{k}_{\perp}}|^2\right).
\end{equation}
This form makes it explicit that the Reynolds and Maxwell contributions to the zonal-flow energy have definite but opposite signs. More specifically, Reynolds stress injects energy into the zonal flow inside the shearing zone, while the Maxwell stress extracts energy from it.

\section{Benchmarks between \texttt{stella}, \texttt{CGYRO} and \texttt{GENE}}
\label{sec:benchmark}
In this appendix, we compare the turbulent heat fluxes calculated using \texttt{stella}, \texttt{CGYRO} and \texttt{GENE} codes.
\subsection{Definition of heat fluxes}
In \texttt{stella}, the turbulent heat flux averaged over the flux-tube volume due to a particular species is defined as
\begin{align}
Q_s & = \frac{1}{V}\int \sd^3\bm{r} \int \sd^3\bm{\bm{v}} \,\frac{m_sv^2}{2} \avg{\left(\bm{v}_{\chi} \cdot \grad x\right) \delta f_s}{\bm{r}} \nonumber \\ & = \frac{1}{V}\int \sd z \,J \frac{c}{B} \unit{b}\cdot \left(\grad y\times \grad x\right)\int\sd x\sd y \int \sd^3 \bm{v}\,\frac{m_sv^2}{2} \avg{\frac{\partial \chi}{\partial y}\delta f_s}{\bm{r}},
\end{align}
where we have written $\int \sd^3 \bm{r}$ in the flux-tube coordinates, and 
\begin{equation}
V = L_x L_y \int \sd z\,J |\grad x|.
\end{equation}
Since \texttt{stella} solves the Fourier-transformed gyrokinetic-Maxwell system of equations, it is more convenient to write $Q_s$ in terms of the Fourier components of the distribution function and fields: 
\begin{equation}
Q_s = \hat{H}\left[\chi_{\bm{k}_{\perp}} \delta f^*_{s,\bm{k}_{\perp}}\right],
\end{equation}
where we have defined the integral operator
\begin{equation}
\hat{H}[\psi] = \frac{1}{\int \sd z\, J |\grad x|}\int \sd z \,J \frac{c}{B} \unit{b}\cdot \left(\grad y\times \grad x\right) \sum_{\bm{k}_{\perp}} ik_y \int \sd^3\bm{v}\,\frac{m_sv^2}{2} \psi,    
\end{equation}
and the Fourier components of the fields
\begin{equation}
\chi_{\bm{k}_{\perp}} = J_0\left(a_s\right)\left(\phi_{\bm{k}_{\perp}} - \frac{\vpar A_{\parallel,\bm{k}_{\perp}}}{c}\right) + \frac{2\mu}{Z_s e}\frac{J_1\left(a_s \right)}{a_s}B_{\parallel,\bm{k}_{\perp}}, \;\; \delta f^*_{s,\bm{k}_{\perp}} = - \frac{Z_s e\phi^*_{\bm{k}_{\perp}}}{T_s}F_s + h^*_{s,\bm{k}_{\perp}}. 
\end{equation}
It will be instructive to extract different contributions to the total heat flux that come from different fields. Thus, the electrostatic heat flux is 
\begin{equation}
\label{eq:esflux}
Q_{s,\mathrm{es}} = \hat{H}\left[J_0(a_s)\phi_{\bm{k}_{\perp}}  \delta f^*_{s,\bm{k}_{\perp}}\right].
\end{equation}
When there is a magnetic perturbation perpendicular to the equilibrium field, parallel heat conduction along the exact field line produces radial heat transport (called `flutter' transport), with the corresponding heat flux given by
\begin{equation}
\label{eq:aparflux}
Q_{s,\apar} = - \frac{1}{c}\hat{H}\left[\vpar J_0(a_s) A_{\parallel,\bm{k}_{\perp}} \delta f^*_{s,\bm{k}_{\perp}}\right].
\end{equation} 
This flutter transport is typically dominated by the electrons due to the large thermal speed at which they travel along field lines. If, in addition, $\bpar \neq 0$, the perturbed field strength can cause a perturbed magnetic drift in the radial direction, which can give rise to a net heat flux
\begin{equation}
\label{eq:bparflux}
Q_{s,\bpar} = \frac{2}{Z_s e}\hat{H} \left[\mu\frac{J_1(a_s)}{a_s}B_{\parallel, \bm{k}_{\perp}} \delta f^*_{s,\bm{k}_{\perp}}\right].
\end{equation}
\subsection{Numerical set-up and simulation results}
\begin{figure}
	\centering
	\begin{tabular}{c} 
		\includegraphics[scale=1.0]{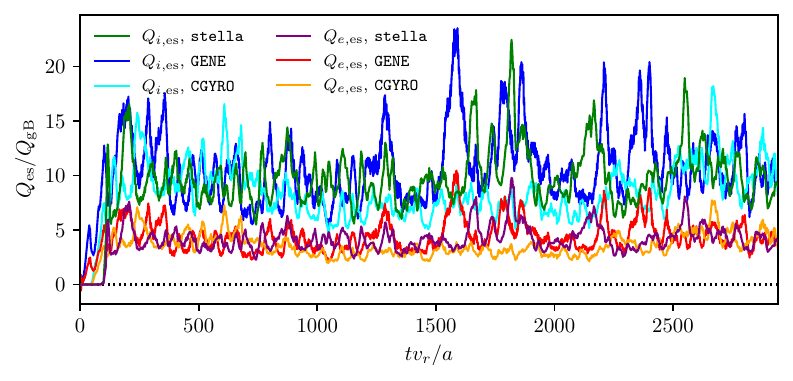} \\
		\includegraphics[scale=1.0]{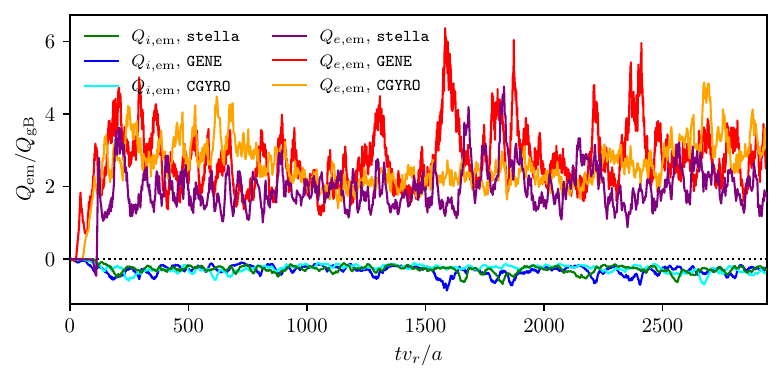}
	\end{tabular}
	\caption{Time traces of heat fluxes normalised to $Q_{\rm{gB}} = n_rT_rv_r\rho^2_r/a^2$. The data are from CBC simulations using \texttt{stella}, \texttt{CGYRO} and \texttt{GENE}. The CBC parameters listed in Table~\ref{tab:norms} are used, with $\beta_e = 0.006$, inclusion of $\bpar$ and a consistent value of $\beta^{\prime}$. Both panels show time traces of different heat flux components, with the top panel the electrostatic heat fluxes and the bottom panel the electromagnetic heat fluxes.}
	\label{fig:heat_flux_benchmark}
\end{figure}
For the CBC benchmark, we choose the set of equilibrium parameters given in Table~\ref{tab:norms}, with the addition of $ \bpar$ and a consistent $\beta^{\prime}$. The value of electron beta is set to $\beta_e=0.006$, and collisions are ignored and replaced by hyper-viscosity. The same wavenumber resolution and box sizes in $x$ and $y$ are used across all three codes. In \texttt{stella}, the box sizes are $L_x = L_y = 126 \rho_r$ in real space and \mbox{$v_{\parallel,\mathrm{max}}/\vths = v_{\perp,\mathrm{max}}/\vths = 3$} in velocity space. The numerical resolutions are $N_x = 171$, $N_y = 22$, $N_z = 32$, $N_{\vpar} = 48$, $N_{\mu} = 12$. In \texttt{CGYRO}, the maximum normalised energy is $\mathrm{E\_MAX}=8$, the number of energy grid points is $\mathrm{N\_ENERGY}=12$, the number of pitch-angle grid points is $\mathrm{N\_XI}=32$, and the number of poloidal grid points is $\mathrm{N\_THETA}=32$. In \texttt{GENE}, the maximum normalised, parallel velocity is $l_v = 3$, the maximum normalised, magnetic moment is $l_w = 9$, and the number of grid points along the equilibrium field line is $n_{z0} = 32$. The velocity-space resolutions are $n_{v0} = 64$ grid points in the parallel velocity and $n_{w0} = 24$ grid points in the magnetic moment.

Fig.~\ref{fig:heat_flux_benchmark} shows the comparison of heat-flux data between the three codes for the electrostatic (\ref{eq:esflux}) (top panel) and electromagnetic (bottom panel) heat fluxes. The latter is the sum of (\ref{eq:aparflux}) and (\ref{eq:bparflux}), $Q_{s,\mathrm{em}} = Q_{s,\apar} + Q_{s,\bpar}$. The time traces show good agreement, on average, between the codes. The time-averaged heat fluxes over a period of saturated state are listed in Table~\ref{tab:averaged_fluxes}. Note that the electromagnetic heat flux is dominated by the electron contribution. Furthermore, the contribution from $Q_{e,\bpar}$, is negligible, so that $Q_{e,\mathrm{em}} \approx Q_{e,\apar}$. For the \texttt{stella} simulation, for example, $Q_{i,\bpar}/Q_{\rm{gB}} = 0.16 \pm 0.04$ and $Q_{e,\bpar}/Q_{\rm{gB}} = 0.17 \pm 0.04 \ll Q_{e,\mathrm{em}}/Q_{\rm{gB}}$. This is not simply a consequence of choosing an artificially small value of $\beta_e$. Indeed, the value adopted here, $\beta_e = 0.006$, lies close to the non-zonal-transition boundary for $q=1.4$ (see Fig.~\ref{fig:q_beta_plots}). These results therefore provide some justification for neglecting $\bpar$ in the CBC simulations presented elsewhere in this paper.

\begin{table}
\centering
\small
\renewcommand{\arraystretch}{1.4}
\begin{tabular}{|c|c|c|c|c|}
\hhline{|-----|}
Heat fluxes$/Q_{\rm{gB}}$ & $Q_{i,\mathrm{es}}$ & $Q_{e,\mathrm{es}}$ & $Q_{i,\mathrm{em}}$ & $Q_{e,\mathrm{em}}$\\
\hhline{|-----|}
\texttt{stella} & $9.9\pm2.5$ & $4.2\pm1.1$ & $-0.28\pm0.10$ & $2.0\pm0.5$ \\
\texttt{CGYRO}  & $8.5\pm2.3$ & $3.6\pm1.0$ & $-0.26\pm0.09$ & $2.6\pm0.6$ \\
\texttt{GENE}   & $11.0\pm3.2$ & $4.5\pm1.4$ & $-0.32\pm0.13$ & $2.8\pm0.8$ \\
\hhline{|-----|}
\end{tabular}
\caption{Heat fluxes averaged over the period from $tv_r/a = 500$ to the end of the simulations for the simulations shown in Fig.~\ref{fig:heat_flux_benchmark}. The errors shown are the root-mean-square deviation from the mean values over the averaging window.}
\label{tab:averaged_fluxes}
\end{table}

\section{Zonal energy transfer}
\label{sec:large_scale}
\subsection{The transfer and zonal flow spectra}
To further justify our focus on the large-scale transfers in Section~\ref{sec:turbulent energy transfers}, we study in detail the energy transfer to the zonal mode due to turbulent stresses at each $k_x$, given by the right-hand side of \eqref{eq:transfer}.

\begin{figure}
  \centering
  \includegraphics[width=0.55\textwidth]{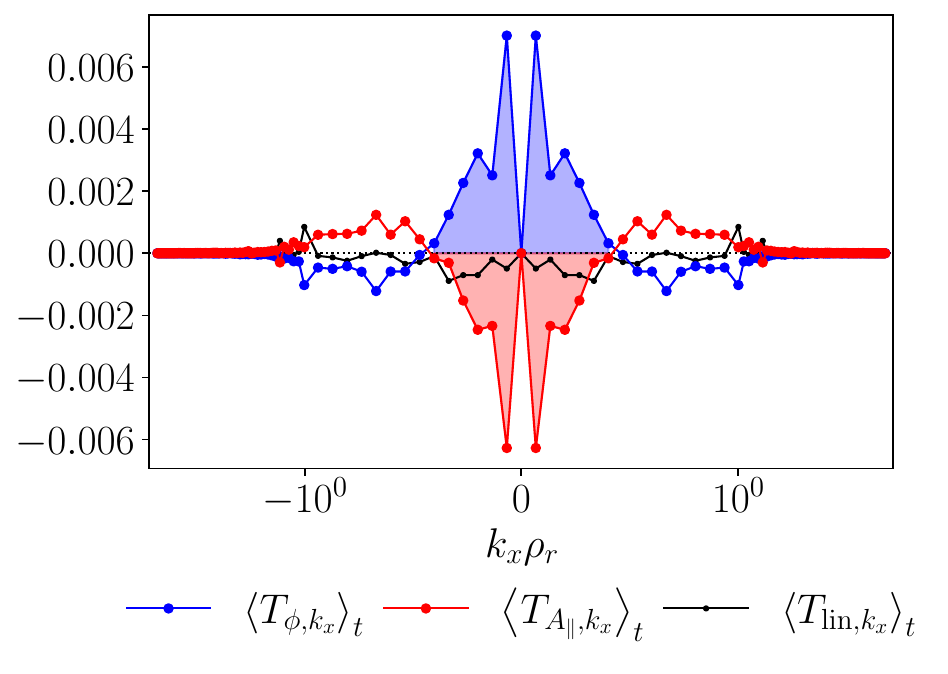}
  \caption[]{Same as the left panel of Fig.~\ref{fig:transfer_spec_1} except here the resolution is \textit{R5} (doubled $k_{x,\mathrm{max}}$ and $k_{y,\mathrm{max}}$ and the same box sizes).}
    \label{fig:transfer_spec_2}
\end{figure}

\begin{figure}
	\centering
	\begin{center}
		\begin{tabular}{ cc } 
			\includegraphics[scale=1.0]{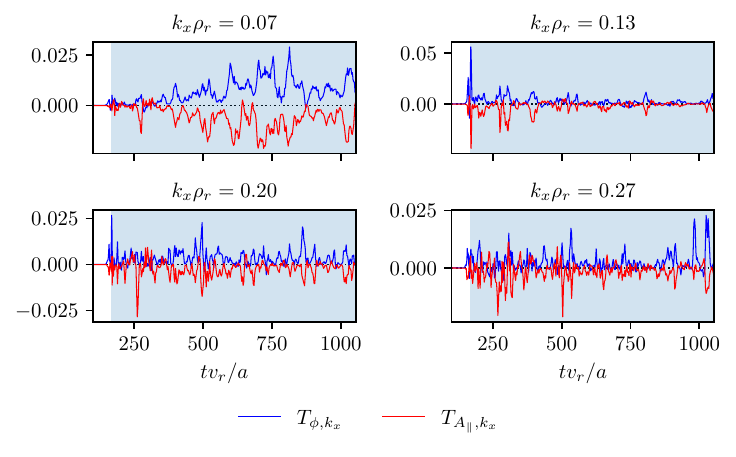}
		\end{tabular}
	\end{center}
	\caption[]{Same as Fig.~\ref{fig:transfer_time_traces} except here the resolution is \textit{R5} (doubled $k_{x,\mathrm{max}}$ and $k_{y,\mathrm{max}}$ and the same box sizes).}
	\label{fig:transfer_time_traces_2}
\end{figure}

In the CBC, the higher-$k_x$ contributions to the transfer spectra shown in the left panel of Fig.~\ref{fig:transfer_spec_1}, including the sharp structures near $k_x \sim 1$, are sensitive to numerical resolution and disappear when the resolution is increased. This can be seen from the transfer spectrum presented in Fig.~\ref{fig:transfer_spec_2}, which is obtained from a simulation with resolution \textit{R5}, where $k_{x,\mathrm{max}}$ and $k_{y,\mathrm{max}}$ are doubled compared to the case shown in the left panel of Fig.~\ref{fig:transfer_spec_1}. It is clear that the higher-$k_x$ contributions to the spectrum have become much less significant. Nevertheless, the time traces of the transfers at four lowest values of $k_x$ in the resolution-\textit{R5} simulation shown in Fig.~\ref{fig:transfer_time_traces_2} appear similar to those from the resolution-\textit{R1} simulation shown in Fig.~\ref{fig:transfer_time_traces}, in terms of the signs and relative sizes of $T_{\phi}$ and $T_{\apar}$. Indeed, the value of $r_{\mathrm{MR}}$ calculated from the \textit{R5} simulation is $0.77$, which differs little from the value $0.74$ calculated for the \textit{R1} simulation.

\subsection{Energy transfer analysis of the toroidal secondary modes}
The large-scale zonal flows discussed so far typically satisfy $k_x \rho_r \lesssim 0.1$. However, as shown by \cite{nies_2024}, in certain regimes, the saturation level of electrostatic ITG turbulence can instead be predominantly controlled by zonal modes at smaller spatial scales, known as toroidal secondary modes (TSMs), with $k_x \rho_r \sim 0.5$. These smaller-scale zonal modes propagate radially due to the radial component of the magnetic drift. The result in \cite{nies_2024} was obtained in the context of electrostatic ITG turbulence with electrons assumed to follow the modified Boltzmann response~\citep{Dorland_1993, Hammett_1993}. What remains to be investigated is whether the inclusion of electromagnetic effects, or simply treating electrons kinetically, alters the characteristic scales of the TSMs. Below, we present two simulations of electromagnetic ITG turbulence where the structure of the TSMs is examined.

\begin{figure}
	\centering
	\begin{center}
		\begin{tabular}{ cc } 
			\includegraphics[scale=0.55]{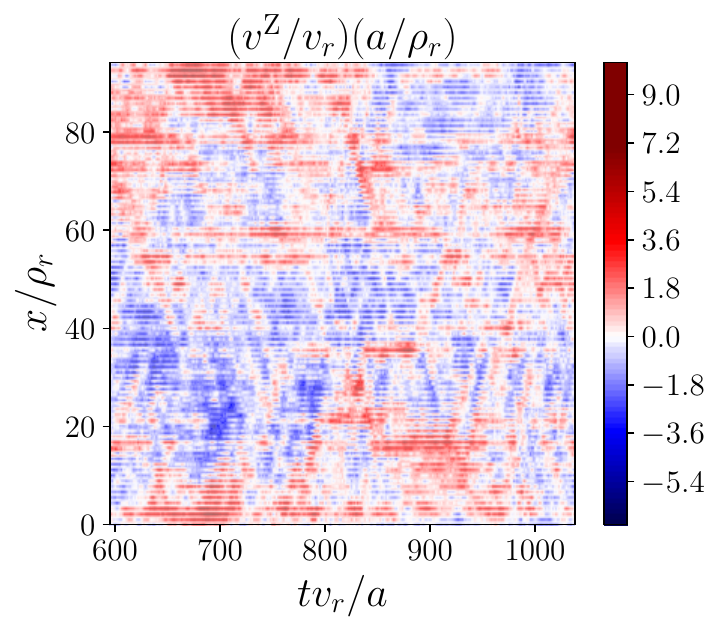} & \includegraphics[scale=0.55]{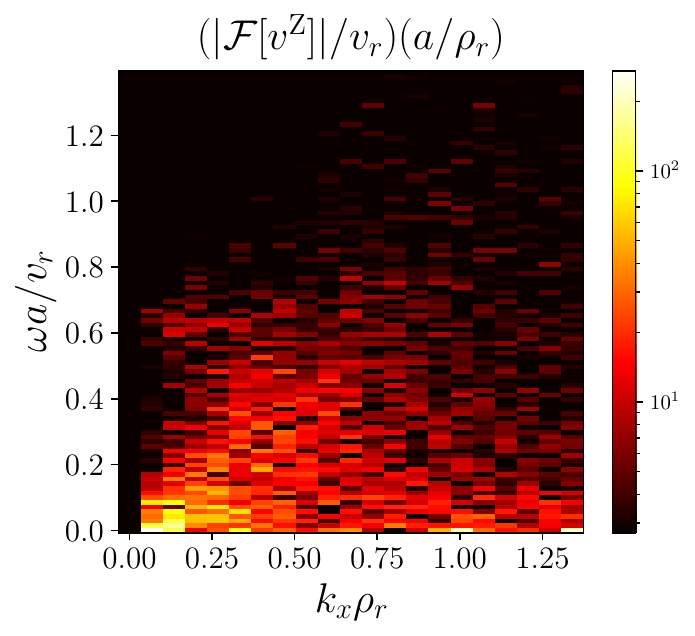}
		\end{tabular}
	\end{center}
	\caption[]{Zonal-flow velocity normalised by $v_r(\rho_r/a)$ obtained from the CBC simulation with lower temperature gradients ($-\sd \ln{T_{i,e}}/\sd \tilde{r} = 2.3$), $q=1.4$, and $\beta_e = 0.0045$. \textbf{Left panel}: Zonal velocity $v^{\rm{Z}}$ versus $x$ and $t$. The presence of propagating, small-scale TSMs can be noticed, though being slightly less pronounced than in Fig.~\ref{fig:zf_xtplot_13.9}. \textbf{Right panel}: Modulus of the Fourier transformed $v^{\rm{Z}}$ versus $k_x$ and $\omega$.}
	\label{fig:zf_xtplot_6.9}
\end{figure}
\begin{figure}
	\centering
	\begin{center}
		\begin{tabular}{ cc } 
			\includegraphics[scale=0.55]{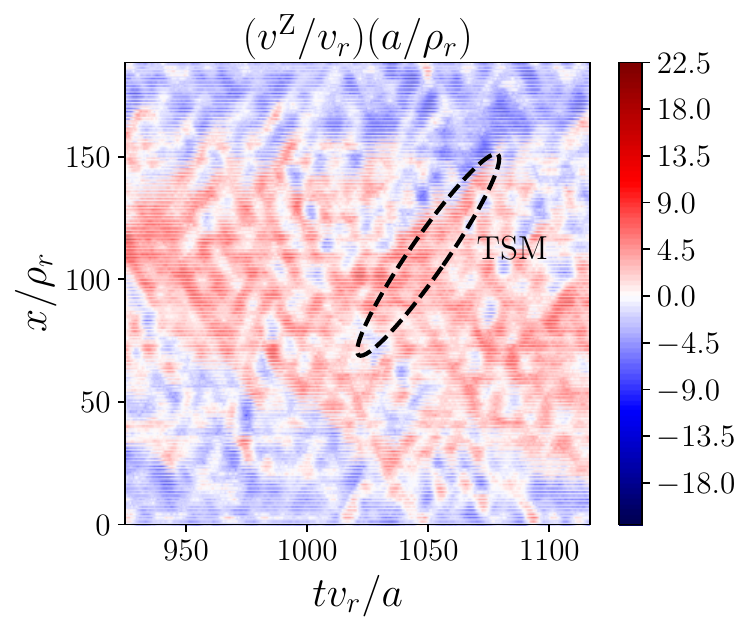} & \includegraphics[scale=0.55]{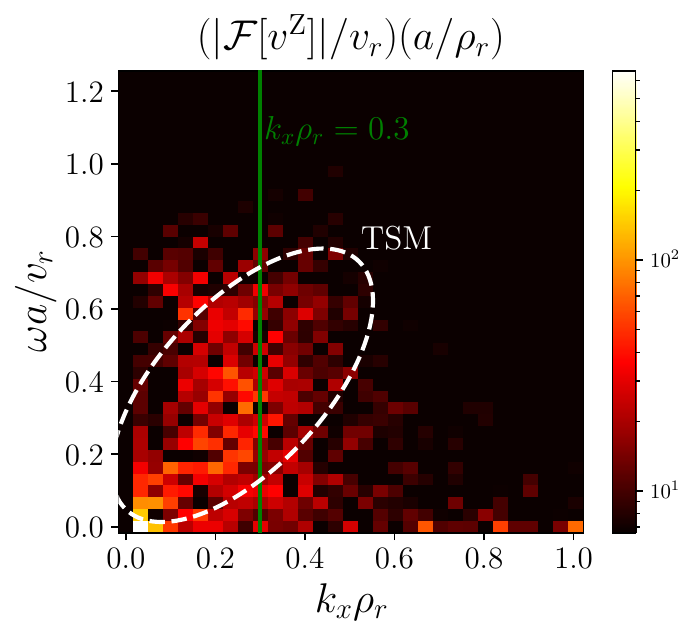}
		\end{tabular}
	\end{center}
	\caption[]{Same as Fig.~\ref{fig:zf_xtplot_6.9} but for a CBC simulation with higher temperature gradients ($-\sd \ln{T_{i,e}}/\sd \tilde{r} = 5.0$), $q=1.4$, and $\beta_e = 0.0015$. Propagating, small-scale TSMs are clearly visible on top of a large-scale background of stationary zonal flows. A green line at $k_x\rho_r = 0.3$ is added for comparison with Fig.~\ref{fig:transfer_omegakx}.}
	\label{fig:zf_xtplot_13.9}
\end{figure}

\begin{figure}
    \centering
    \includegraphics[width=0.5\linewidth]{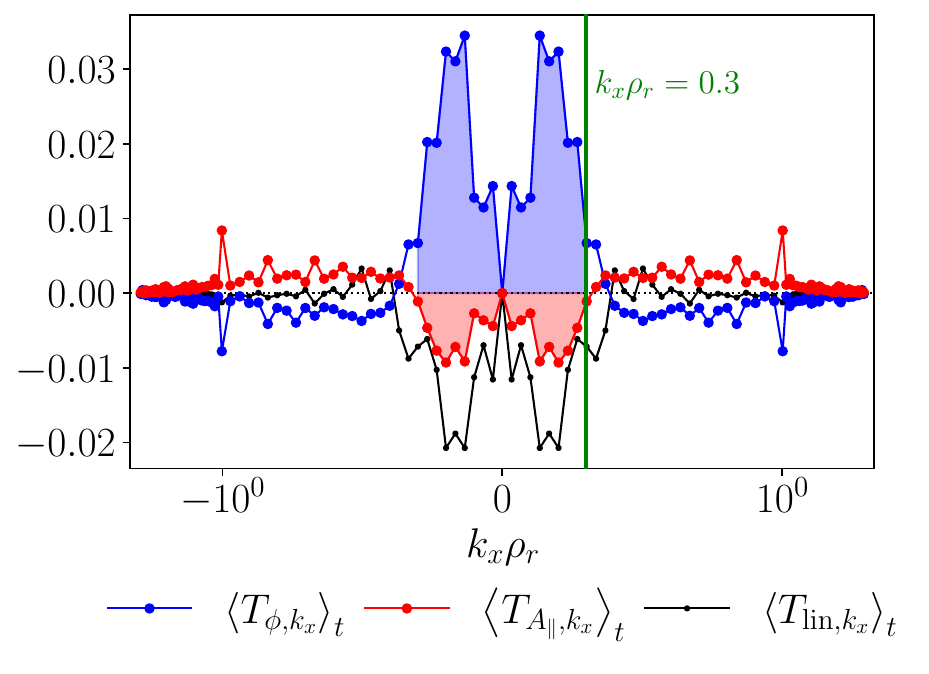}
    \caption{The $k_x$-spectrum of time-averaged energy transfers normalised by $n_rT_rv_r\rho^2_r/a^3$ for the simulation shown in Fig.~\ref{fig:zf_xtplot_13.9}. The blue and red regions show the large-scale ($|k_x\rho_r|\leq  0.3$, a location marked by a green line, same as in Fig.~\ref{fig:zf_xtplot_13.9}) transfers $T_{\phi, k_x}$ and $T_{\apar, k_x}$, respectively. As seen from the right panel of Fig.~\ref{fig:zf_xtplot_13.9}, the green line marks the approximate location of the TSM cut-off.}
    \label{fig:transfer_omegakx}
\end{figure}

The first of these simulations uses the CBC parameters from Table~\ref{tab:norms}, resolution \textit{R1}, $q=1.4$, $\beta_e = 0.0045$, and $D_{\mathrm{hyper}} = 0.2$. The second uses a similar set of parameters, except with $\tilde{R} = 2.78$, $\beta_e = 0.0015$, $-\sd \ln(n_{i,e})/\sd \tilde{r} = 0.8$, $-\sd \ln(T_{i,e})/\sd \tilde{r} = 5.0$, and $D_{\mathrm{hyper}} = 1.0$, and resolutions $N_{\vpar} = 48$, $N_{\mu} = 12$, $L_x = L_y = 188\rho_r$, $k_{x,\mathrm{max}}\rho_r = 2.1$, $k_{y,\mathrm{max}}\rho_r = 1.4$, and $N_z = 32$. The major difference is that the temperature gradients are much higher in the second simulation. Note that the second set of parameters are only used in the current appendix. 

The resulting zonal structures in the saturated turbulence are shown in Figs.~\ref{fig:zf_xtplot_6.9} and~\ref{fig:zf_xtplot_13.9}. In each figure, the left panel shows the zonal velocity $v^{\rm{Z}}$ versus $(x,\,t)$, while the right panel shows the modulus of its Fourier transform versus $(k_x,\,\omega)$. Compared to the lower-temperature-gradient case in Fig.~\ref{fig:zf_xtplot_6.9}, the higher-temperature-gradient case in Fig.~\ref{fig:zf_xtplot_13.9} exhibits clearer signatures of TSMs. In particular, the left panel of Fig.~\ref{fig:zf_xtplot_13.9} shows small-scale TSMs propagating radially on top of a large-scale background of stationary zonal flows. In the electromagnetic simulations considered here, however, the stationary zonal flows dominate in amplitude over the TSMs, in contrast to the electrostatic cases with Boltzmann electrons reported in \cite{nies_2024}, where the TSMs were found to be dominant. This difference is likely due to the inclusion of kinetic electrons.

The appearance of TSMs can be further confirmed by the right panel of Fig.~\ref{fig:zf_xtplot_13.9}, where, even though the dominant peak is located at low $k_x$ and $\omega = 0$ (stationary zonal flows), a propagating branch with phase speed $\omega/k_x \approx (3/2)\,\rho_r v_r/a \approx 4\rho_r v_r/R$, which is of the same order of magnitude as the estimate $\omega/k_x \sim 2\rho_r v_r/R$ reported in \cite{nies_2024}, is clearly visible. These propagating TSMs appear to be of the kind that \cite{nies_2024} identified as playing a crucial role in setting the saturation level of electrostatic ITG turbulence, described by the `grand-critical-balance' conjecture \citep{Ghim_2013}. 

In Fig.~\ref{fig:transfer_omegakx}, to examine whether the large-scale transfer defined by~(\ref{eq:large_scale_transfer}) captures the contribution of the TSMs, we present the transfer spectrum for the simulation shown in Fig.~\ref{fig:zf_xtplot_13.9}. As shown by the spectrum, both the Reynolds and Maxwell transfers are dominated by low-$k_x$ contributions and decrease to nearly zero around $k_x \rho_r \approx 0.3$. As seen from the right panel of Fig.~\ref{fig:zf_xtplot_13.9}, this wavenumber also corresponds approximately to the cut-off scale for the TSMs. Therefore, at least for the CBC cases considered here, the large-scale transfer prescription (\ref{eq:large_scale_transfer}) should remain valid for predicting non-zonal transitions, even when TSMs are clearly visible in the overall zonal structure.

\section{Long-term evolution and transport hysteresis}
\label{sec:long_evolution_and_transport_hysteresis}
\subsection{Evolution in larger boxes}
\label{sec:long_evolution}
In this appendix, we show that increasing the box size may lead to a long-term growth of the box-size zonal mode, which gradually suppresses the turbulent transport and the energy transfer by the Maxwell stress out of the zonal flows. This is consistent with the observations made by \cite{Rath_2022}. We shall clarify how one calculates the transfer ratio $r_{\mathrm{MR}}$ in such cases.

\begin{figure} 
\centering 
\begin{center} 
\begin{tabular}{ cc } 
\includegraphics[scale=0.7]{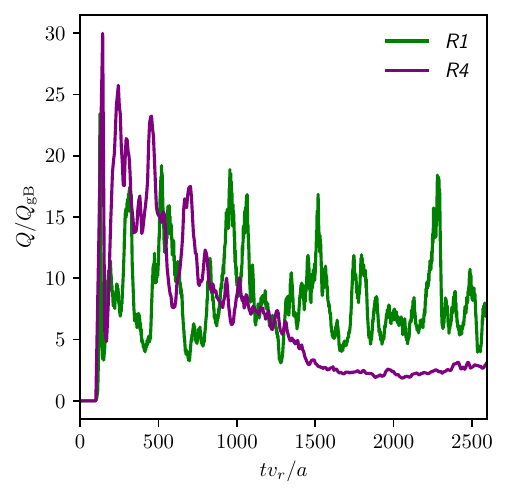} & \includegraphics[scale=0.7]{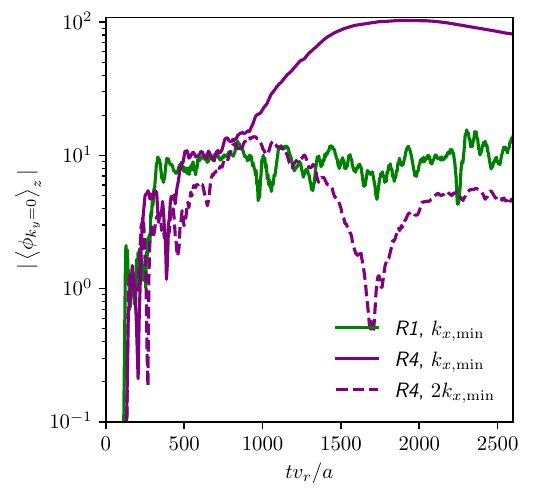} 
\end{tabular}
\end{center} 
\caption[]{\textbf{Left panel}: Time traces of the total heat flux normalised by $Q_{\rm{gB}} = n_r T_r v_r \rho_r^2 / a^2$ for two CBC simulations with $q=1.4$ and $\beta_e = 0.007$ and resolutions \textit{R1} (green) and \textit{R4} (purple). \textbf{Right panel}: Time traces of the zonal modes $|\avg{\phi_{k_y=0}}{z}|$ normalised by $T_r \rho_r / e a$, for the same simulations, shown again in solid green (\textit{R1}) and solid purple (\textit{R4}). The second-lowest-$k_x$ zonal mode from the \textit{R4} simulation \,(\textit{R4}, $2k_{x,\mathrm{min}}$) -- which has the same $k_x$ as the box-size mode in the \textit{R1} simulation \,(\textit{R1}, $k_{x,\mathrm{min}}$) -- is shown as a dashed purple line.} 
\label{fig:total_heat_and_zonal_modes} 
\end{figure}

In the left panel of Fig.~\ref{fig:total_heat_and_zonal_modes}, we show the time trace of the heat flux for the CBC with $q = 1.4$ and $\beta_e = 0.007$ at two different resolutions, \textit{R1} and \textit{R4}, the latter with double the box size of the former in both $x$ and $y$ directions.  In the \textit{R1} simulation, the heat flux reaches a steady level shortly after saturation. In contrast, the \textit{R4} simulation requires a significantly longer time to settle and ultimately saturates at a lower level. This delayed saturation is caused by the slow, long-term growth of the box-size zonal mode in the \textit{R4} case. The right panel of Fig.~\ref{fig:total_heat_and_zonal_modes} shows the corresponding time traces of the zonal electrostatic potential. Compared with the \textit{R1} case, the box-size zonal mode in the \textit{R4} simulation continues to grow over a much longer timescale and eventually reaches a much larger amplitude, leading to a stronger suppression of turbulence and hence a reduced heat flux.

\begin{figure}
    \centering
    \includegraphics[scale=1.0]{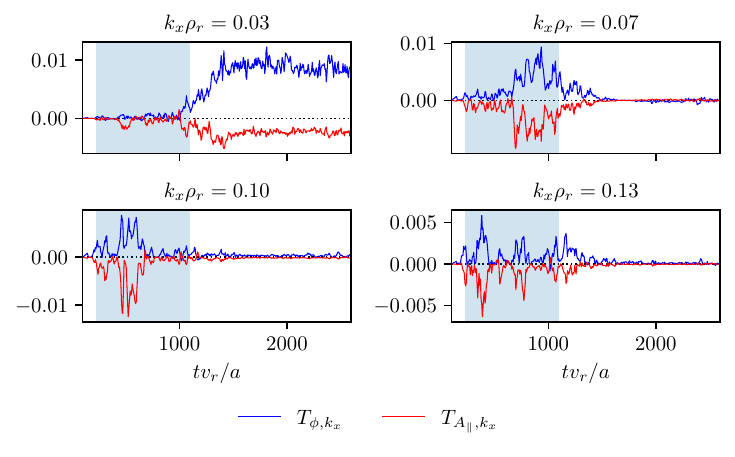}
    \caption{Same as Fig.~\ref{fig:transfer_time_traces} except here the resolution is \textit{R4} (doubled box size in both directions but the same $k_{x,\mathrm{max}}$ and $k_{y,\mathrm{max}}$).}
    \label{fig:transfer_time_traces_3}
\end{figure}

\begin{figure}
    \centering
    \includegraphics[scale=0.5]{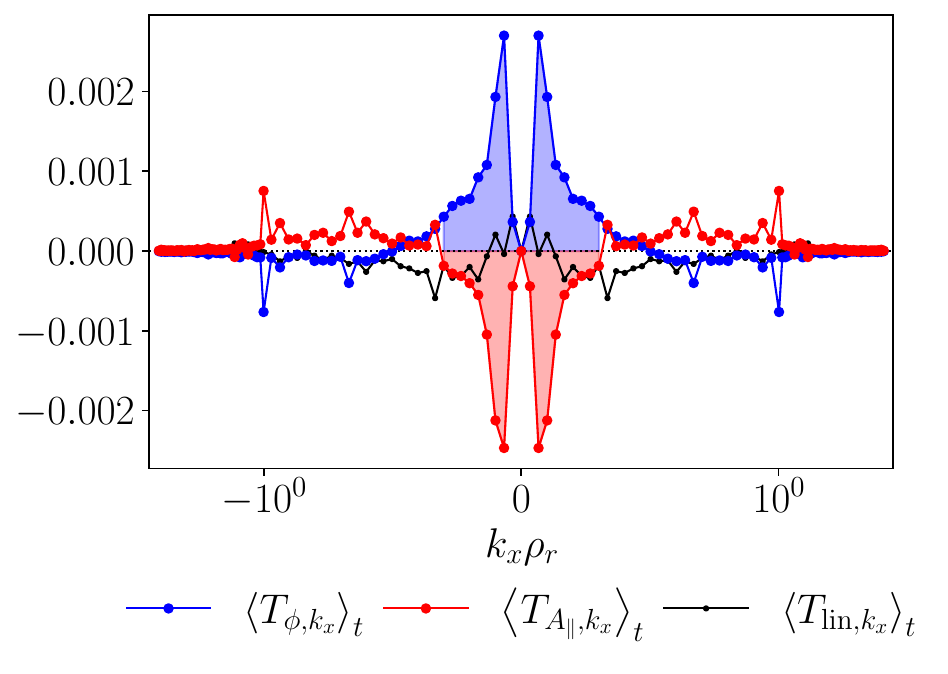}
    \caption[]{Same as the left panel of Fig.~\ref{fig:transfer_spec_1} except here the resolution is \textit{R4} (doubled box size in both directions but the same $k_{x,\mathrm{max}}$ and $k_{y,\mathrm{max}}$).}
    \label{fig:transfer_spec_3}
\end{figure}

For the \textit{R4} simulation, we also present the time traces of the energy transfers for four values of $k_x$ in Fig.~\ref{fig:transfer_time_traces_3} and the time-averaged transfer spectrum in Fig.~\ref{fig:transfer_spec_3}. There are a few notable observations. First, higher-$k_x$ contributions to the transfer spectrum now appear less important compared to those of the \textit{R1} simulation (left panel of Fig.~\ref{fig:transfer_spec_1}), and most of the transfer activity is concentrated at large scales, where the Reynolds and Maxwell stresses compete. Secondly, the transfers are dominated by the box-size zonal mode in the long term, resulting in an overall reduction in the Maxwell-to-Reynolds transfer ratio after an initial period of saturation up to $tv_r/a \approx 1100$. This type of behaviour was also found by \cite{Rath_2022}. There, a positive feedback loop that maintains strong zonal flows was proposed: the weakening of the Maxwell stress over time allows the zonal flow to grow stronger, and the resulting enhancement of the box-size zonal flow further suppresses the Maxwell stress. This positive feedback loop suggests that electromagnetic fluctuations are suppressed by the large-scale zonal flows more effectively than their electrostatic counterparts. As a result, saturation might be achieved even for cases slightly above the transition threshold -- provided the system is restarted from a state in which the box-size zonal flow has already developed to a level where the Maxwell stress becomes subdominant relative to the Reynolds stress. Further discussion of this can be found in Appendix~\ref{sec:hysteresis}.

At first glance, the results presented in this appendix may suggest that box-size convergence has not been fully achieved. However, the precise values of the heat fluxes are not of primary importance for the present discussion. The essential question is whether the large-scale transfer data provide sufficient information to determine how far a given case lies from the non-zonal transition threshold. It is worth noting in this context that CBC simulations above the transition, e.g., with $q=1.4$ and $\beta_e = 0.01$, fail to saturate, regardless of whether resolution \textit{R1}, \textit{R4}, or \textit{R5} is used, provided that the initial condition is random noise with small amplitudes. This raises the possibility that a shared signature in the transfer data across these simulations may serve as a robust indicator of the distance to the threshold. Consider the CBC simulation with $q=1.4$ and $\beta_e = 0.007$, which lies just below the threshold. Both \textit{R1} and \textit{R5} simulations produce similar time traces of the transfer functions (Fig.~\ref{fig:transfer_time_traces} and Fig.~\ref{fig:transfer_time_traces_2}, respectively), with $T_{\phi}$ and $T_{\apar}$ staying comparable. The corresponding time trace of the \textit{R4} simulation (Fig.~\ref{fig:transfer_time_traces_3}) appears different in the long term, but matches the \textit{R1} and \textit{R5} behaviour over the interval where the box-size zonal mode remains relatively low in amplitude. Thus, for the \textit{R4} simulation, the value of $r_{\mathrm{MR}}$ calculated from the transient period (indicated by the shading in Fig.~\ref{fig:transfer_time_traces_3}) is $r_{\mathrm{MR}} \approx 0.84$, which agrees well with the values obtained from the \textit{R1} ($r_{\mathrm{MR}} \approx0.74$) and \textit{R5} ($r_{\mathrm{MR}} \approx0.77$) simulations. 

The key practical conclusion is that, when using energy-transfer data exhibiting long-term evolution to assess the distance from the non-zonal transition threshold, one should compute the transfer ratio $r_{\mathrm{MR}}$ using a time average taken over the transient period before the long-term growth of the zonal modes. Evaluating $r_{\mathrm{MR}}$ over the transient period is also physically motivated, as this is the period during which it is decided whether the turbulence saturates -- if the system enters a non-zonal state during the transient, owing to a strong $T_{\apar}$ that drains energy from the zonal flows, those flows never reach amplitudes large enough relative to the turbulence to trigger the positive feedback loop required to maintain themselves.

\subsection{Transport hysteresis}
\label{sec:hysteresis}
In Appendix~\ref{sec:long_evolution}, we showed that in a simulation with a larger box size in both $x$ and $y$, the box-size zonal mode continued to grow over long times and eventually suppressed the energy transfer out of zonal flows due to Maxwell stress. \cite{Rath_2022} demonstrated that if such simulations were restarted during a phase where the Maxwell transfer was weak -- even with a restart value of $\beta_e$ above the non-zonal threshold obtained from simulations initialised with small noise, as we did for all cases in Fig.\ref{fig:q_beta_plots} -- the turbulence could still saturate. Therefore, the turbulence appeared to have access to two different saturated states depending on the initial condition -- a transport hysteresis.

\begin{figure}
	\centering
	\begin{center}
		\begin{tabular}{ cc } 
			\includegraphics[scale=0.7]{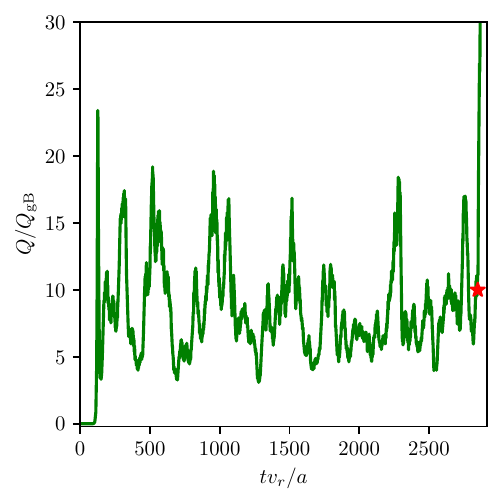} & \includegraphics[scale=0.7]{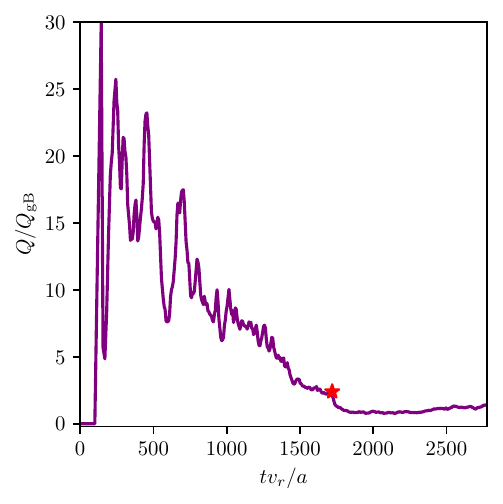}
		\end{tabular}
	\end{center}
	\caption[]{Time traces of the total heat flux, normalised by $Q_{\rm{gB}} = n_r T_r v_r \rho_r^2 / a^2$, for two CBC simulations with $q=1.4$ and, initially, $\beta_e = 0.007$, at two resolutions: \textit{R1} (smaller box, left panel) and \textit{R4} (larger box, right panel). The simulations are restarted at the times marked by the red stars with the value of $\beta_e$ increased to $0.01$, above the transition threshold, which is at $\beta_e \approx 0.008$ (see Fig.~\ref{fig:q_beta_plots}).}
	\label{fig:total_heat_time_trace2}
\end{figure}

Here, we explore this possibility by restarting our simulations of the CBC with $q = 1.4$ with initial conditions designed to access a saturated state with finite heat flux where otherwise an extreme value would be obtained. Fig.~\ref{fig:total_heat_time_trace2} shows the time traces of the total heat flux for two such simulations. Both runs begin with $\beta_e = 0.007$ and are later restarted with a higher value, $\beta_e = 0.01$, at the times indicated by the red stars. The left panel shows a resolution-\textit{R1} run, which produces a diverging heat flux following the restart. In contrast, the right panel shows that, for a resolution-\textit{R4} run (larger box), the heat flux remains saturated for a considerably longer interval and persists in a quasi-steady state. We think that this difference is due to the fact that the box-size zonal mode remains substantially weaker in the \textit{R1} run than in the \textit{R4} run, as shown in the right panel of Fig.~\ref{fig:total_heat_and_zonal_modes}. As a result, at the restart time, the Maxwell transfer continues to compete strongly with the Reynolds transfer in the \textit{R1} run, so increasing $\beta_e$ further immediately leads to erosion of the zonal flows, and hence to a high-transport state. In contrast, in the \textit{R4} run, the energy transfer from the zonal flows by the Maxwell stress is already much weaker at the restart time than the energy transfer into the zonal flows by the Reynolds stress, due to the suppression of the former by the box-size zonal mode, providing a buffer that prevents a moderate increase in the Maxwell transfer with $\beta_e$ from destroying the zonal flows.

As stated previously in Appendix~\ref{sec:long_evolution}, if the \textit{R4} simulation is instead initialised with noise, the heat flux diverges, indicating that turbulence saturation depends on the initial condition, particularly the zonal structure. Thus, our numerical experiments confirm the existence of the transport hysteresis reported by \cite{Rath_2022}.

\end{document}